\documentclass{article}
\usepackage{arxiv}
\usepackage{enumitem}

\usepackage{xcolor}
\usepackage{url}

\usepackage{adjustbox}
\usepackage{graphicx} 
\usepackage[linesnumbered,ruled,vlined]{algorithm2e}
\usepackage{caption}
\usepackage{listings}
\usepackage{xcolor}
\usepackage{multirow}
\usepackage{colortbl}
\usepackage{booktabs}
\usepackage{hhline}
\usepackage{subfigure}
\usepackage{appendix}
\usepackage{pifont}
\usepackage{tikz}

\usepackage{framed}

\usepackage{tcolorbox}

\usepackage[affil-it]{authblk}

\newcommand{\tab}{\hspace*{1em}}
\newcommand{\code}[1]{{\fontfamily{cmtt}\fontseries{m}\fontshape{n}\selectfont\small{#1}}}

\newcommand{\sysname}{ECMO\xspace}
\newcommand{\sysnamenospace}{ECMO}

\newcommand{\ballnumber}[1]{\tikz[baseline=(myanchor.base)] \node[circle,fill=.,inner sep=1pt] (myanchor) {\color{-.}\bfseries\footnotesize #1};}

\definecolor{dkgreen}{rgb}{0,0.6,0}
\definecolor{gray}{rgb}{0.5,0.5,0.5}
\definecolor{mauve}{rgb}{0.58,0,0.82}

\definecolor{mygreen}{RGB}{0 ,205, 102}
\begin{document}

\title{\sysnamenospace: Peripheral Transplantation to Rehost Embedded Linux Kernels}

	\author[1,2]{Muhui Jiang}
	\author[1]{Lin Ma}
	\author[1]{Yajin Zhou\thanks{Corresponding author (yajin\_zhou@zju.edu.cn).}\hspace*{0.4em}}
	\author[1]{Qiang Liu}
	\author[3]{Cen Zhang}
	\author[4]{Zhi Wang}
	\author[2]{Xiapu Luo}
	\author[1]{Lei Wu}
	\author[1]{Kui Ren}
	\affil[1]{Zhejiang University}
	\affil[2]{The Hong Kong Polytechnic University}
	\affil[3]{Nanyang Technological University}
	\affil[4]{Florida State University}

\maketitle


\begin{abstract}

Dynamic analysis based on the full-system emulator
QEMU is widely used for various purposes.
However, it is challenging to run
firmware images of embedded devices in QEMU,
especially the process to boot the Linux kernel
(we call this process  rehosting the Linux kernel
in this paper).
That's because embedded devices usually use different
system-on-chips (SoCs)
from multiple vendors and only a limited
number of SoCs are currently supported in QEMU.

In this work, we propose a technique
called \textit{peripheral transplantation}.
The main idea is to transplant the device drivers of
designated peripherals into the Linux kernel binary.
By doing so, it can replace the peripherals in the kernel
that are currently unsupported in QEMU with
supported ones, thus making the Linux kernel \textit{rehostable}.
After that, various applications can be built.

We implemented this technique inside a prototype system
called \sysname and applied it to $815$ firmware images,
which consist of $20$ kernel versions and 37 device models. The result
shows that \sysname can  successfully transplant peripherals for all the $815$ Linux kernels.
Among them, $710$ kernels can be successfully rehosted, i.e.,
launching a user-space shell ($87.1\%$ success rate). 
The failed cases are mainly because the  root
file system format (\textit{ramfs}) is not supported by the kernel. Meanwhile, we are able to inject rather complex drivers (i.e., NIC driver) for all the rehosted Linux kernels by installing kernel modules.
We further build three applications, i.e., kernel crash analysis,
rootkit forensic analysis, and kernel fuzzing, based on the rehosted kernels
to demonstrate the usage scenarios of  \sysname.

\end{abstract}

\section{Introduction}

IoT devices (or embedded devices) are becoming popular~\cite{iot_market}, 
many of which run Linux-based operating systems~\cite{UsenixSec14_Costin}.
At the same time, hundreds of vulnerabilities are discovered every year for the Linux kernel~\cite{cve_kernel}.
Once the devices are compromised, attackers can control them to launch further attacks. 
As such, the security of embedded devices, especially the kernel, deserves a thorough analysis. 

Dynamic analysis has been widely used for various purposes~\cite{rept2020atc,cui2018rept,harrison2020partemu,maier2019unicorefuzz,zheng2019firmafl,jiang2007stealthy}.
It can monitor the runtime behavior of the target system, complementing the
static analysis~\cite{redini2020karonte,shoshitaishvili2015firmalice,UsenixSec14_Costin, hernandez2017firmusb}.
Rehosting, also known as emulation, is used to run a target system
inside an emulated environment, e.g., QEMU, and provides the capability
to introspect the runtime state.
Based on this capability, different applications, e.g.,
kernel crash analysis, rootkit forensic analysis, and kernel fuzzing, can be built upon.
Running the Linux kernel in QEMU for the desktop system is a solved
problem. However, embedded systems usually have different
system-on-chips (SoCs) with customized hardware peripherals from multiple vendors. Due to the diverse peripherals in the wild, it is not practical for QEMU to support all kinds of peripherals in any SoC, which is tedious and error-prone.
In this case, embedded Linux kernels, which depend on the unsupported peripherals, may stuck, halt, or crash during the rehosting process.
Thus,
how to rehost the embedded Linux kernels in QEMU is still an open research question.

Previous research~\cite{chen2016firmadyne,kim2020firmae} provides the capability
of rehosting user-space programs by running a customized Linux kernel
for a single type of SoC that is already supported in QEMU.
This works well because user-space programs
mainly depend on standard system calls that are provided
by the underlying Linux kernel.
Different from user-space programs,  the OS kernel (e.g., Linux kernel) interact with hardware peripherals that are usually different in different SoCs.

Some researchers have proposed to use real devices to perform the dynamic
analysis~\cite{zaddach2014avatar,avatar2,eric2019pretender,talebi2018charm}.
Such solutions do not scale since there exist a large number
of embedded devices. Other researchers work towards
the bare-metal systems~\cite{feng2019p2im,dice,clements2019hal}, i.e., embedded
systems without an OS kernel or having a thin layer of abstraction.
However, the methodology cannot be directly used to rehost the Linux kernel as the Linux kernel is far more complicated than the bare-metal ones.

\smallskip
\noindent \textbf{Our Approach}\tab
In this work, we propose a solution called \textit{peripheral transplantation. It is device-independent, and works towards the Linux kernel binary without the
need of the source code of the target system}.
The main idea is, instead of manually adding emulation support of various
peripherals in QEMU, we can transplant the device drivers of
designated peripherals into the target Linux kernel binary.
It replaces the peripherals in the target Linux kernel
that are currently unsupported in QEMU with
supported ones, thus making the Linux kernel
\textit{rehostable}.
In particular, given a Linux kernel retrieved from the firmware image
of an embedded device, our system turns it into a rehostable
one that can be successfully booted in QEMU. After that,
various applications can be built
to analyze the rehosted kernel.

\begin{figure}[t]
	\centering
	\includegraphics[width=0.5\linewidth]{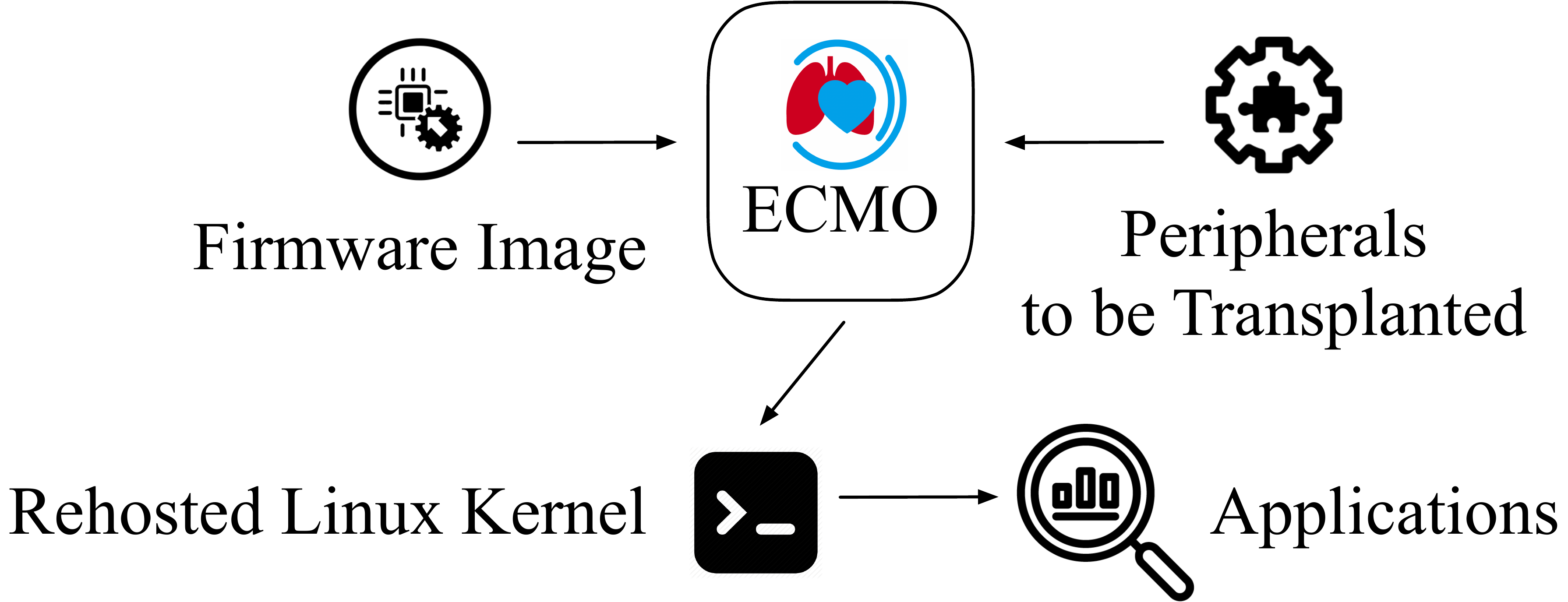}
	\caption{The overview of our system (\sysname)}
	\label{fig:ecmo_intro}
	
\end{figure}

Specifically, our system transplants two components, i.e., the emulated models
of peripheral into QEMU and their device drivers into the Linux kernel.
Transplanting a peripheral model requires us to build the
hardware emulation code for the specified (or simplified) peripheral
and integrate it into QEMU. This is straightforward 
Since QEMU has provided us with APIs to add new peripheral models.

However, transplanting a driver into the Linux kernel binary
is non-trivial since we need to achieve three
design goals. \emph{First}, we need to substitute the original (unsupported) device driver with the transplanted one.  Since the peripheral driver
is initialized with indirect calls in Linux kernel, we need to
locate functions and rewrite the function pointers in a stripped binary
on the fly, which is challenging.
\emph{Second}, the transplanted driver should not affect the memory
view of the original kernel.
Otherwise, the memory holding the transplanted driver can be
overwritten since the Linux kernel is not aware of
the existence of that memory region.
\emph{Third}, the transplanted driver needs to invoke APIs in the
Linux kernel. Otherwise, the transplanted driver cannot
function as desired.

To address these challenges, we implement and integrate
the \textit{peripheral transplantation} technique into
QEMU to create a prototype called \sysname.
We detail the system design and implementation in
Section~\ref{sec:design}.
Figure~\ref{fig:ecmo_intro} shows the overview of \sysname. It
receives the firmware image and the peripherals to be transplanted.
Then it transplants the peripherals to the Linux kernel binary
to make it rehostable in QEMU and launch a shell.
Then we can build various applications on it.

We apply our system on $815$ Linux kernel binaries extracted from
firmware images, including $20$ different kernel versions,
$37$ device models, and $24$ vendors. 
Our experiment shows that \sysname can successfully transplant peripherals
for all $815$ Linux kernels. Among them, $710$ are able to  boot and launch a shell.
The failed cases are due to the unsupported root
file system format (\textit{ramfs}) in the rehosted kernel.
To demonstrate the functionality and usefulness of our system, we build and port three
applications, including kernel crash analysis, rootkit forensic analysis, and kernel fuzzing. Note that, the applications
themselves
are not the contribution of our work. They are used to demonstrate the usage scenarios
of our system. Other applications that can be built on QEMU can also be ported. 

In summary, this work makes the following main contributions.

	\noindent \textbf{Novel technique}\tab We propose a \textit{device-independent} technique
	called \textit{peripheral transplantation} that can rehost Linux kernels
	of embedded devices without the availability of the source code.
	
	\noindent \textbf{New system}\tab We implement and integrate the \textit{peripheral transplantation} technique into QEMU, to create a prototype system called \sysname.
	
	\noindent \textbf{Comprehensive evaluation}\tab We apply \sysname to Linux kernels from $815$ different images. It can transplant peripherals for all the Linux kernels and successfully launch the shell for $710$ ones.

To engage with the community, we will release the source code of our system and the
dataset of firmware images.
We have provided a docker image~\cite{ecmo_docker} that contains the
\sysname system for  testing.

 \section{Background}

\subsection{Linux Kernel}
\label{subsec:kernel}

Linux kernel source code can be categorized into three types according to their functionalities.
The first type is the \textit{architecture independent code}, which contains the
core functionality used by all CPU architectures. 
The second type is \textit{architecture dependent code}.
For instance, the sub-directories under the \textit{arch/} directory
contain the code for multiple CPU architectures.
The third type is \textit{board-specific code}, which is used by specific board (machine).
For instance, the directory  \textit{arch/arm/versatile/} contains the code used by the machine named \textit{versatile}.
The kernel compiled for one machine usually cannot be directly booted
on other machines (or QEMU instances that emulate different machines.)

\begin{figure}[t]
	\centering
	\begin{lstlisting}
		MACHINE_START(VERSATILE_AB, "ARM-Versatile AB")
		    .atag_offset = 0x100,
		    .map_io	= versatile_map_io,
		    .init_early	= versatile_init_early,
		    .init_irq = versatile_init_irq,
		    .init_time = versatile_timer_init,
		    .init_machine = versatile_init,
		    .restart = versatile_restart,
		MACHINE_END
	\end{lstlisting}
	\caption{The machine description for \textit{ARM-Versatile AB}.}
	\label{fig:machine_desc}
\end{figure}

\subsection{ARM Machines}
\label{sec:arm_machine}
Embedded systems usually use SoCs from
multiple vendors with different designs.
For instance, they contain different  peripherals.
Each SoC is expressed as a machine in the Linux kernel.
Manufacturers develop the \textit{board support package} (BSP)
(e.g., drivers of peripherals) so that Linux kernel can use these peripherals.

Linux kernel introduces the structure \textit{machine\_desc} for ARM to describe
different machines. The structure \textit{machine\_desc} provides interfaces
to implement BSPs. For example, Figure~\ref{fig:machine_desc}
shows an example of one machine \textit{ARM-Versatile AB} in the Linux kernel (Version 3.18.20).
It initializes function pointers and data pointers with its implementation. 
Specifically, in line 5, the function pointer \textit{init\_irq} is assigned the value as 
\textit{versatile\_init\_irq}. During the booting process, the Linux kernel will invoke
the function \textit{machine\_desc$\rightarrow$init\_irq} to initialize the IC (interrupt controller). 
The same logic applies to the function pointer \textit{init\_time}. Linux kernel
invokes the function \textit{machine\_desc$\rightarrow$init\_time} to initialize the timer.

\begin{figure}[t]
	\centering
	\begin{lstlisting}
    //UART read call back
    static uint64_t serial_mm_read(void *opaque, 
        hwaddr addr, unsigned size) {
	    SerialMM *s = SERIAL_MM(opaque);
	    return serial_ioport_read(&s->serial, 
                             addr >> s->regshift, 1);
    }
    //register read/write call back functions
    static const MemoryRegionOps serial_mm_ops = {
    	.read = serial_mm_read,
	    .write = serial_mm_write,
	    ...
    };
	\end{lstlisting}
\caption{The callback functions for UART emulation in QEMU}
\label{fig:qemu_uart}
\end{figure}

\subsection{QEMU}
QEMU~\cite{qemu} is one of the most popular full-system emulators.
It emulates different machines by providing
different machine models. A machine model consists of
CPU, memory, and different kinds of peripheral models.
To emulate a peripheral, QEMU registers the read/write callback functions
for the MMIO (memory-mapped I/O) address space of the peripheral.
Once the Linux kernel running inside QEMU reads from or writes into
the address inside the MMIO range, the registered
callback functions inside QEMU
will be invoked to emulate the peripheral. Basically, it maintains an internal
state machine to implement the peripheral's functionality. Figure~\ref{fig:qemu_uart}
shows an example of the registered callback functions for UART emulation.
Specifically, when the Linux kernel reads from the MMIO space of the
emulated UART device (e.g., \code{0x01C42000}), the
\code{serial\_mm\_read} function will be invoked by QEMU to emulate the read access.

\section{Challenges and Our Solution}
\label{sec:design}

The main goal of our work is to rehost
Linux kernel binaries that are originally running on
embedded systems in QEMU.
This lays the foundation of applications that rely on the
capability to introspect runtime states of the Linux kernel,
e.g., kernel crash and vulnerability analysis~\cite{rept2020atc,cui2018rept},
rootkit forensic analysis~\cite{wang2008countering,riley2008guest},
and kernel fuzzing~\cite{maier2019unicorefuzz,schumilo2017kafl}. 

\subsection{Challenges}
Rehosting the Linux kernel on QEMU
faces the following challenges.

\smallskip
\noindent \textbf{Peripheral dependency}\tab
Rehosting the Linux kernel requires QEMU
to emulate the peripherals, e.g.,
the interrupt controller, that the Linux kernel depends on.
During the booting process, Linux kernel will read from or write into the peripheral registers and execute the code according to the state specified by the value of peripheral registers.
Without the emulation of these peripherals,
the rehosted kernel will halt or crash during the booting process.

\smallskip
\noindent \textbf{Peripheral diversity}\tab
SoCs vary widely~\cite{openwrtsoc} and different vendors, e.g., Broadcom, Marvell may design and develop different SoCs. These new
SoCs introduce many new peripherals that are not currently supported
in QEMU and the open-sourced mainstream of the Linux kernel.
Due to the diversity of peripherals, there are still a large number of devices that are not supported. Meanwhile, manually developing peripheral emulation
routine is tedious and error-prone, especially due to the diversity of peripherals.
Thus, the diversity of
peripherals brings significant challenge to build a general emulator, which can re-host various Linux kernels of embedded
devices.

\smallskip
\noindent \textbf{Lack of public information}\tab
The information (e.g., specifications, datasheets, and source code)
of SoCs and firmware images are usually not public. This is because vendors may not release the detailed hardware specification. Furthermore,  vendors may not release the source code immediately after releasing the image and not all vendors strictly follow the GPL license~\cite{german2009empirical,duan2017identifying}. Meanwhile, the binary of the Linux kernel
is stripped and has no particular headers (i.e., ELF section
headers) or debugging information. 
These 
obstruct the diagnosis of failures when
adding emulation support of new SoCs in QEMU.



\subsection{Our Solution: Peripheral Transplantation}
\label{sec:solution}
In this work, we propose a technique called
\textit{peripheral transplantation}. The main idea is, instead
of manually adding emulation support of
various peripherals in QEMU, we can \textit{replace the
peripherals that are used in target Linux kernels with
existing peripherals in QEMU.}
By doing so,
we can rehost the Linux kernel and the kernel functionality is intact (Section~\ref{sec:functionality}). 

\begin{figure}[t]
	\centering
	\includegraphics[width=0.8\linewidth]{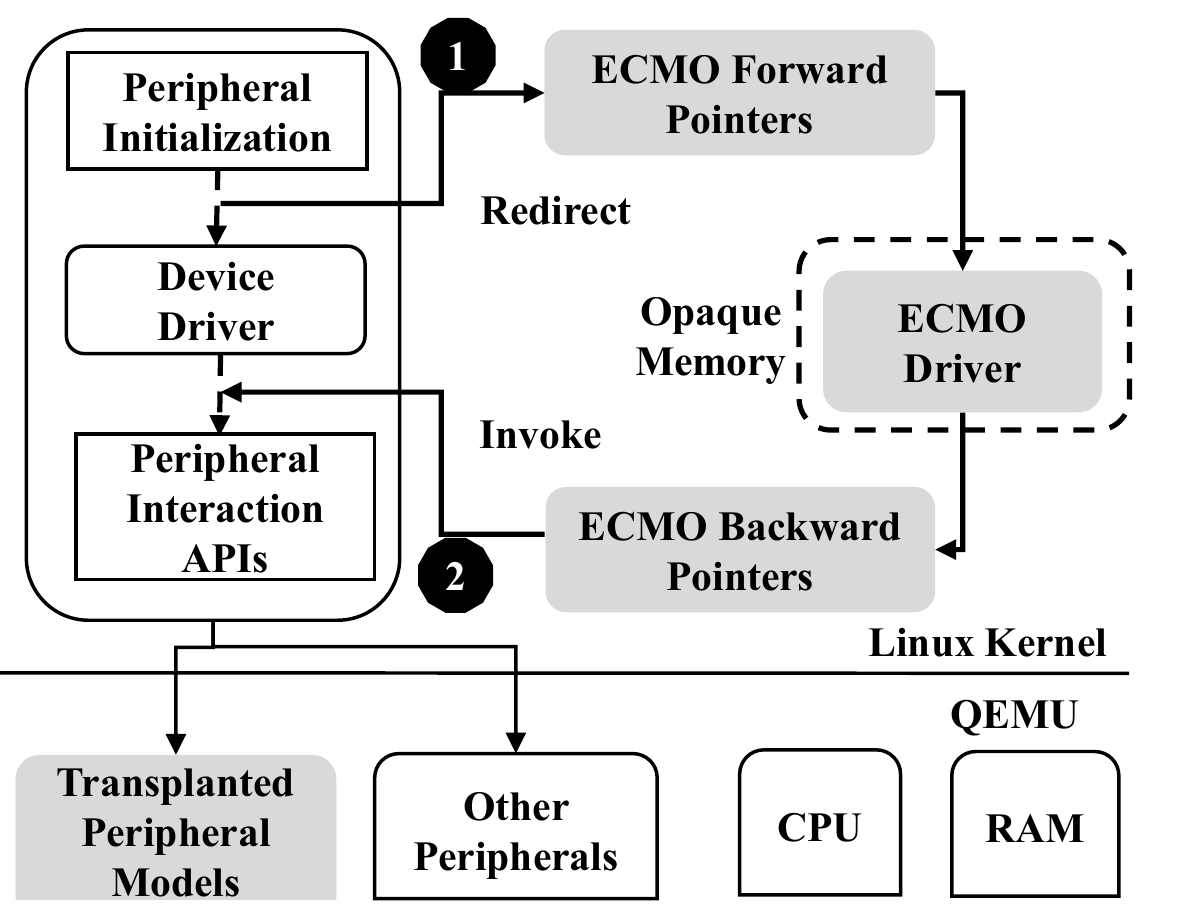}
	
	\caption{The overview of peripheral transplantation.}
	\label{fig:overall_new}
	
\end{figure}

\begin{figure*}[t]
	\centering
	\includegraphics[width=0.85\linewidth]{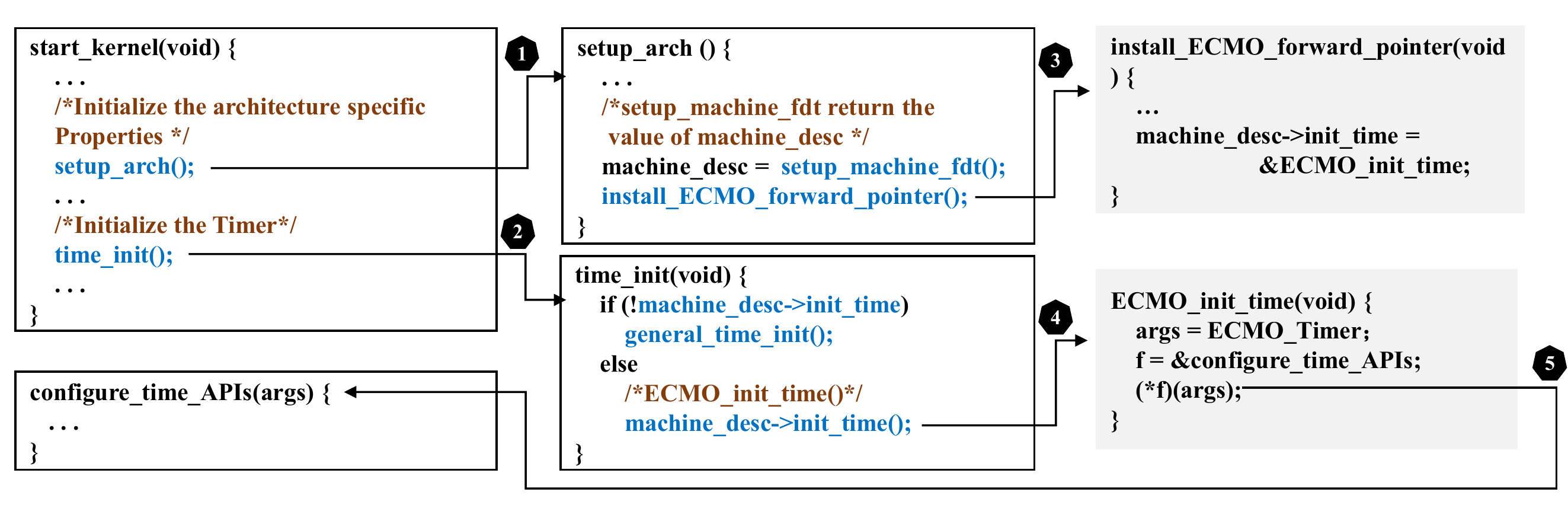}

	\caption{A concrete example of peripheral transplantation.}
	\label{fig:example_new}
	
\end{figure*}

Figure~\ref{fig:overall_new} shows the overview of
peripheral transplantation. This involves
the injection of peripheral models
into QEMU
and the \textit{ECMO Driver} into the Linux kernel. To distinguish them from original ones of the (emulated) machine, we call the transplanted peripheral models \textit{ECMO Peripheral}.
To let the kernel use the transplanted
\textit{ECMO Driver}, our system identities
the functions that are used to initialize
device drivers (\textit{ECMO Forward Pointers})
and redirects them to the functions inside the \textit{ECMO Driver} (Fig.~\ref{fig:overall_new} ~\ballnumber{1}). 
Moreover, our system identifies the APIs
that are responsible for interacting with
peripheral models.
These APIs are used by the \textit{ECMO Driver} to
communicate with the transplanted peripheral models (Fig.~\ref{fig:overall_new} ~\ballnumber{2}).
The addresses of these functions are called
\textit{ECMO Backward Pointers} in this paper. We will elaborate how to identify the ECMO Pointers in Section~\ref{sec:identify_ecmo_pointers}.

Note that, to ensure the \textit{ECMO Driver} does not affect
the memory view of the rehosted Linux kernel, we propose the
concept of the \textit{opaque memory}. This memory region is available on
the emulated machine but cannot be seen by the Linux kernel.
As such, we can prevent the kernel from allocating
memory pages that are reserved for the \textit{ECMO Driver}.
We will elaborate this in Section~\ref{sec:generate_driver}.

\begin{figure}[t]
	\centering
	\includegraphics[width=\linewidth]{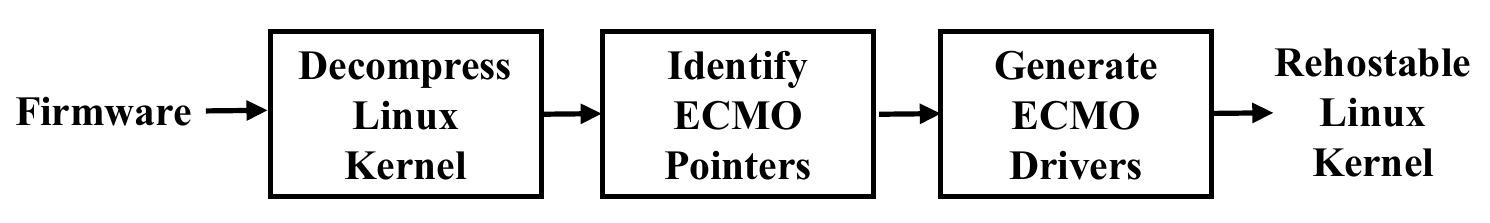}
	
	\caption{The work flow of our system.}
	\label{fig:flow_new}
	
\end{figure}

\subsection{An Illustration Example of Peripheral Transplantation}

Fig.~\ref{fig:example_new} shows a concrete example of transplanting one peripheral (i.e., timer) into the Linux kernel.
In particular, the function \code{start\_kernel} is responsible
for initializing the Linux kernel. It will invoke several different functions, including \code{setup\_arch} and and \code{time\_init}.

The function \code{setup\_arch} will setup architecture-related configurations
and initialize the \code{machine\_desc} structure (Fig.~\ref{fig:example_new}~\ballnumber{1}). This structure
contains multiple function pointers (\textit{ECMO Forward Pointers}) that will be used to
initialize corresponding drivers. Our system first locates the function 
\code{setup\_arch} and then injects a function
(\code{install\_ECMO\_forward\_pointers}) to change the pointers to our
own ones (Fig.~\ref{fig:example_new}~\ballnumber{3}).

When the function \code{init\_time} is invoked to initialize the timer (Fig.~\ref{fig:example_new}~\ballnumber{2}),
the \code{ECMO\_init\_time}, which is pointed by \code{machine\_desc->}
\code{init\_time}, will be invoked to initialize the injected timer driver
(\textit{ECMO Driver}) in QEMU (Fig.~\ref{fig:example_new}~\ballnumber{4}) (through \textit{ECMO Forward Pointers}), instead of the original one. Accordingly, this function
will invoke APIs (through \textit{ECMO Backward Pointers}) in the Linux kernel to
interact with the \textit{ECMO Peripheral} (Fig.~\ref{fig:example_new}~\ballnumber{5}).

Note that, the code snippets in Fig.~\ref{fig:example_new} are for
the illustration purpose.
\textit{Our system does not rely on the availability of the source code. It directly works towards
the Linux kernel binary  that is retrieved from a firmware image.}

 \section{System Design and Implementation}
\label{sec:design}

In order to rehost Linux kernels, our system first extracts and decompresses the
Linux kernel from the given firmware image (Section~\ref{sec:decompress}).
We then apply multiple
strategies to identify both \textit{ECMO Forward and Backward Pointers} (Section~\ref{sec:identify_ecmo_pointers}).
These pointers are essential for \textit{ECMO Drivers}. At last, we
semi-automatically generate \textit{ECMO Drivers} and load them at runtime to
boot the kernels (Section~\ref{sec:generate_driver}).
Fig.~\ref{fig:flow_new} shows the overall workflow.

\subsection{Decompress Linux Kernel}
\label{sec:decompress}

\begin{figure}[t]
	\centering
	\begin{lstlisting}
	  Assembly code:
      mov	r0, #0
	    str	r0, [r2], #4	
	    str	r0, [r2], #4
	    str	r0, [r2], #4
	    str	r0, [r2], #4
	    cmp	r2, r3
	    blo	1b
	    tst	r4, #1
	    bic	r4, r4, #1
	    blne	cache_on
      mov	r0, r4 //r0 stores the value of output_start
	    mov	r1, sp			
	    add	r2, sp, #0x10000
	    mov	r3, r7
	    bl	decompress_kernel
	    // we can dump the decompressed Linux kernel after 
	    // function decompress_kernel returns
	
	  Simplified C code:	    
	    void decompress_kernel(uint32 output_start, args) 
	\end{lstlisting}
	\caption{The assembly code that invokes function \textit{decompress\_kernel}, which is in \textit{arch/arm/boot/compressed/head.S}.}
	\label{fig:decompress_kernel}
	
\end{figure}

Firmware image usually consists of the OS, which is the Linux kernel, and user applications. However, the Linux kernel inside the firmware images is usually in the compressed zImage format. 
To identify ECMO Pointers inside the Linux kernel, we need to first extract the Linux kernel and decompress it. With the decompressed Linux kernel, we can utilize different strategies to locate the ECMO Pointers.

Specifically, we feed the firmware image to firmware extraction tool (i.e., Binwalk) to extract the kernel image. Then we directly feed the 
extracted kernel image (with added
u-boot information) to QEMU. Since the code inside the zImage
to decompress the Linux kernel does not operate on the peripherals
(except the UART to show the message of decompressing Linux kernel),
it can be successfully executed in vanilla QEMU. 

As shown in Fig.~\ref{fig:decompress_kernel}, 
function
\code{decompress\_kernel} is used to decompress the kernel.
Its first parameter (i.e., \code{output\_start}) indicates the start address of the decompressed kernel.
Thus, if we can identify when \code{decompress\_kernel} is invoked,
we can get the first parameter by checking the machine register (\code{R0} in ARM) and dump the decompressed Linux kernel.

We notice that the function \textit{decompress\_kernel} is invoked  by the assembly code in \textit{arch/arm/boot/compressed/head.S}. We observe that this snippet of assembly code remains unchanged in different kernel versions. With this observation, we identify the address of instruction \code{BL decompress\_kernel} by strictly comparing the
execution trace of QEMU and the hard coded assembly code. After finding the instruction, we can obtain the address of the function \textit{decompress\_kernel} and the value of \textit{output\_start} according to the execution trace. With this information, we can dump the decompressed Linux kernel after the function \textit{decompress\_kernel} returns.

By doing so, we can automatically
retrieve decompressed Linux kernels from firmware images. 

\subsection{Identity ECMO Pointers}
\label{sec:identify_ecmo_pointers}
Our system needs to obtain the addresses of two essential types
of functions in the Linux kernel.
Specifically, the \textit{ECMO Forward Pointers} contain the functions
that are used by the Linux kernel to initialize device drivers.
We dynamically hook and redirect them to \textit{ECMO Drivers} at runtime
in QEMU. The \textit{ECMO Backward Pointers} contain the APIs that are used
by the \textit{ECMO Driver} to invoke functions provided by the Linux kernel to
interact with emulated peripherals in QEMU. 

However, precisely identifying ECMO Pointers is not easy.
The main challenge is the decompressed Linux kernel is stripped and
only contains the binary data.
It has neither meaningful headers nor debugging symbols and
contains thousands of functions. Furthermore, the Linux kernel
is compiled with different compilers and compiling options,
which can result in different binaries. Thus, we cannot have any assumption
on the compiling options or compilers. We also cannot rely on run-time symbol tables like \code{/proc/kallsym} because they are only available \emph{after} booting. To address these problems, we propose following strategies that can automatically identify ECMO Pointers.

The basic idea to identify ECMO Pointers is leveraging the
source code of the mainline Linux kernel,
which provides valuable information for functions that need
to be located. Note ECMO Pointers are functions in \textit{architecture independent code} or \textit{architecture dependent code} (Section~\ref{subsec:kernel}),
which is in mainline Linux kernel and open-source.
For instance, if we find that a function uses
a specific string by reading the source code, then we can easily locate
this function inside the binary by locating the function that has
references to the same string. Of course, this simple strategy may
not always work, since some functions do not have such obvious patterns
or multiple functions can refer to the same string.
In our work, we take three different strategies to
locate ECMO Pointers (Section~\ref{sec:pointer_strategy}).
We illustrate each step to identify ECMO
Pointers in the following.

\subsubsection{Disassemble the Linux Kernel}

The first step is to disassemble the Linux kernel for further analysis,
including constructing the control flow graph and identifying function
boundaries. Accurately disassembling the ARM binaries is still
challenging, especially when the binary is stripped~\cite{jiang2020empirical}.
In our work, we choose to ensure that this step does not introduce false negatives,
i.e., all the code sections should be dissembled. Otherwise, we cannot identify
the functions if they are not correctly disassembled. 
However, we can tolerate the false positives, i.e., the inline data may be wrongly
disassembled as code. The strategies described in Section~\ref{sec:pointer_strategy}
can help us to filter out these false positives.

After disassembling the Linux kernel and constructing the control flow graphs, we further locate
function boundaries by combining the algorithm introduced in Nucleus ~\cite{andriesse2017compiler} and angr~\cite{angr}. Necleus can identify the functions indirectly called while angr locates the function according to the prologue. These two tools can help to reduce the false negatives and guarantee that the required function addresses (ECMO Pointers) will be located during the disassembly process. Finally, we  build a mapping for each
function and various types of information, e.g., number of basic blocks,
string references, number of called functions and etc. This mapping describes
the signature (or portrait) of each function. Note that, our system does not
require that the constructed control flow graph is sound or complete, as long as
they can provide enough information for further analysis (Section~\ref{sec:pointer_strategy}).

\subsubsection{Locate Pointer Addresses}
\label{sec:pointer_strategy}

\begin{algorithm}[!ht]
	\footnotesize
	\SetKwInput{KwInput}{Input}                
	\SetKwInput{KwOutput}{Output}              
	\DontPrintSemicolon
	\KwInput{The decompressed Linux kernel $LKB$; \\
		The source code of ECMO Pointers $SC$ (architecture independent code or architecture dependent code);}
	\KwOutput{The addresses of ECMO Pointers $FA$;}
	
	\SetKwFunction{FIdentify}{Identify}
	\SetKwFunction{FGenerateFunctions}{GenerateFunctions}
	\SetKwFunction{FDisassembly}{Disassembly}
	
	\SetKwProg{Fn}{Function}{:}{}
	\Fn{\FIdentify{$LKB$,$SC$}}{
		$CFG$ = Disassembly($LKB$)\;
		$Generated\_Functions$ = GenerateFunctions($CFG$)\;
		
		\For(){$S\_F$ in $SC$}{
			\For(){$G\_F$ in $Generated\_Functions$}{
				\For(){Filtering\_Strategy in Filtering\_strategies}{
					\If(){Filtering\_Strategy($S\_F$,$G\_F$)}{
						Append $G\_F$ to $S\_F.Candidates$
					}
					
				}
			}
		}
		
		\For(){$S\_F$ in $SC$}{
			\If(){Length($S\_F.Candidates$) == 1}{
			    $FA[S\_F]$ = $S\_F.Candidates$
			}
		}
		\Return $FA$\;
	}
	\;
 	
	\caption{The algorithm to identify the addresses of ECMO pointers from the Linux kernel binary.}
	\label{alg:match_new}
\end{algorithm}

\begin{figure}[t]
	\centering
	
	\subfigure[Specific constant string: the constant string is referenced by a data pointer (i.e., foo\_offset+0x200).
	 ]{\includegraphics[width=0.48\textwidth]{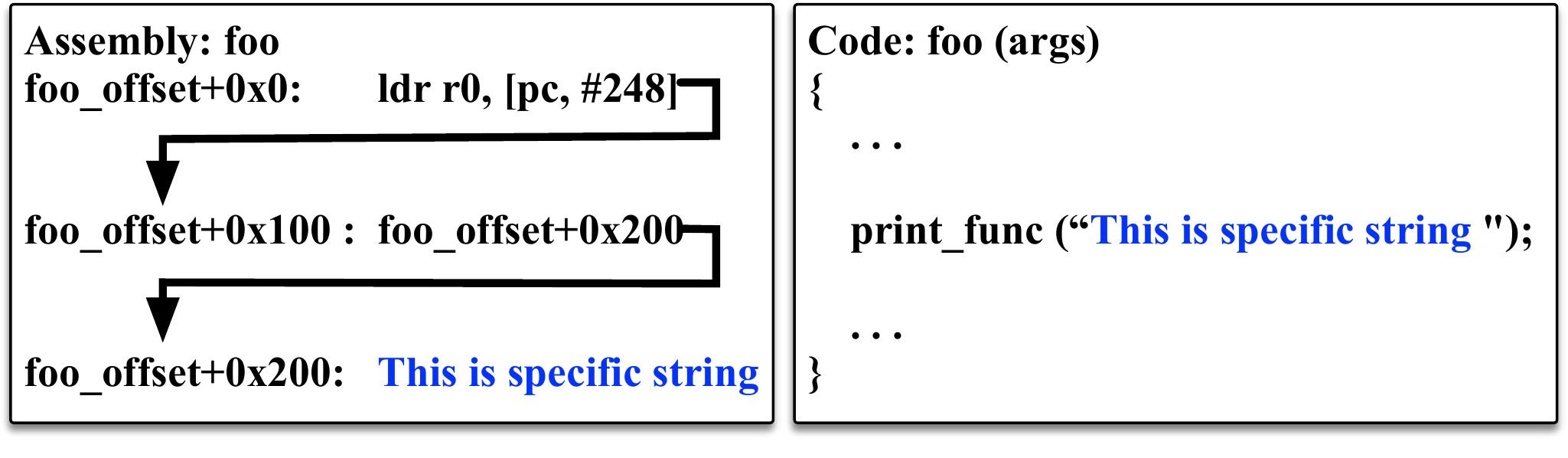}\label{fig:strategy_1_1}}
	
	\subfigure[Warning information: line number (i.e., 386) is the operand of assembly code; file name (i.e., /path/to/source.c) is a constant string. ]{\includegraphics[width=0.48\textwidth]{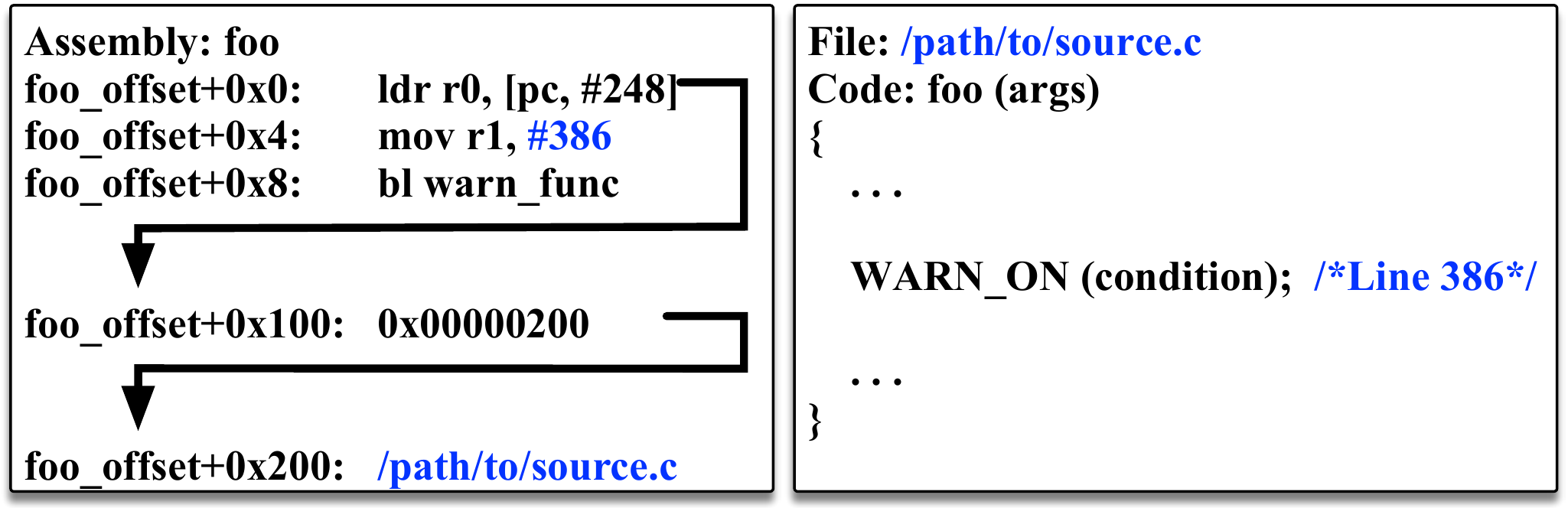}\label{fig:strategy_1_2}}

	\caption{Strategy-I: Lexical information}
	\label{fig:strategy_1}

\end{figure}

Algorithm~\ref{alg:match_new} describes the process to locate
pointer addresses of ECMO Pointers in the decompressed
Linux kernel binary, i.e., \textit{LKB}. Note that, we first need to get the
source code of the functions, i.e., \textit{SC}, inside the mainline Linux kernel.
The outputs of this algorithm are the addresses of ECMO Pointers in
the kernel binary, i.e., \textit{FA} (line 12).

First, we disassemble the decompressed Linux kernel, construct
the control flow graph (line 2) and generate function boundaries (line 3). 
Then for the source code function of each ECMO Pointer (line 4), we loop through the generated functions (line 5) and apply different filtering strategies
(line 6). If one filtering strategy can
identify one address as a candidate address of the ECMO Pointer (line 7),
this address will be appended to the candidate list (line 8). 
Finally, we check the candidates of each ECMO Pointer (line 9). If there is only one
candidate (line 10), it means the address of this ECMO Pointer is  successfully identified in the kernel binary (line 11).

\smallskip
\noindent \textbf{Strategy-I: Lexical information}\tab 
The first strategy uses the lexical information inside a function
as its signature, e.g., a specific constant string and the warning
information. If the function we want to identify has such
strings, we can then lookup the disassembly code to find the functions
that have data references to the same string. The line number and
file name in the warning information can further help to locate the function.

Fig.~\ref{fig:strategy_1_1} shows a pair of the disassembled code and
the source code in the mainline Linux kernel.
In the source code, the function \code{foo} contains
a specific constant string ``\textit{This is a specific string}".
In the assembly code, the instruction at \code{foo\_offset+0x0} will load the data
pointers (i.e., \code{foo\_offset+0x100}) using the \code{LDR} instruction.
The data pointer refers to another pointer (i.e.,  \code{foo\_offset+0x200}),
which contains the same constant string.
Based on this, we can locate function \code{foo}
in the disassembled kernel. Fig.~\ref{fig:strategy_1_2} shows a
similar example with the warning information. The WARN\_ON will call function \textit{warn\_func}. The first parameter is the filename, which is a specific constant string. The second parameter is the line number of WARN\_ON. Usually, the line number is hard coded as an operand of instruction after compilation. Thus, functions containing specific constant string or warning information can be easily identified.

\begin{figure}[t]
	\centering
	\subfigure[Caller relationship: Required\_foo is the caller of Identified\_foo]{\includegraphics[width=0.47\textwidth]{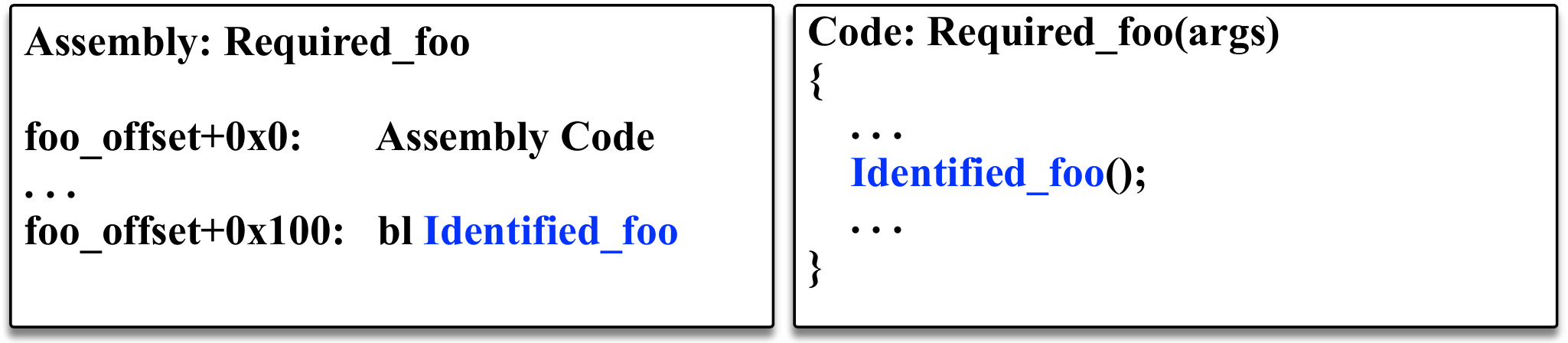}\label{fig:strategy_2_1}}
	
	\subfigure[Callee relationship: Required\_foo is the callee of Identified\_foo ]{\includegraphics[width=0.47\textwidth]{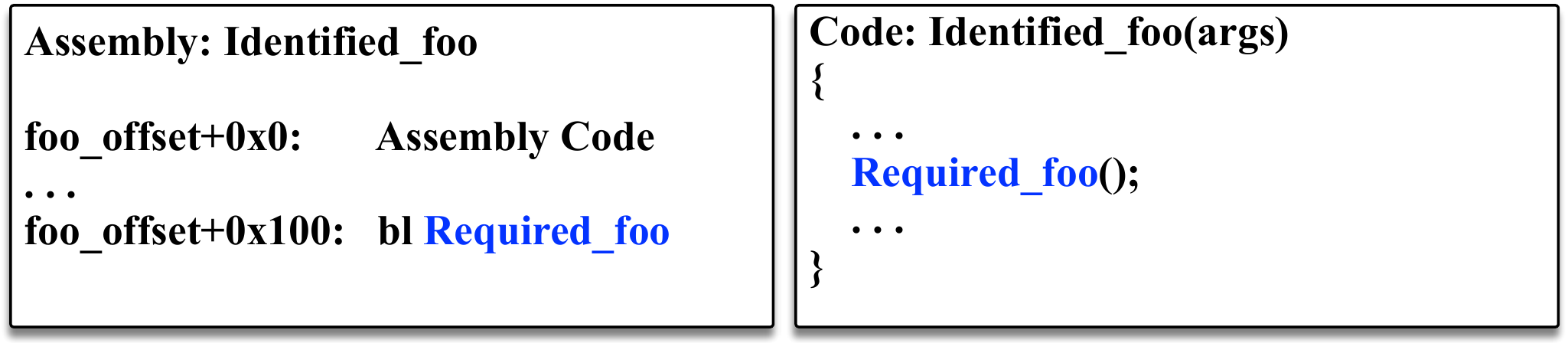}\label{fig:strategy_2_2}}
	
	\subfigure[Sibling relationship: Required\_foo and Identified\_foo are both called by foo  ]{\includegraphics[width=0.47\textwidth]{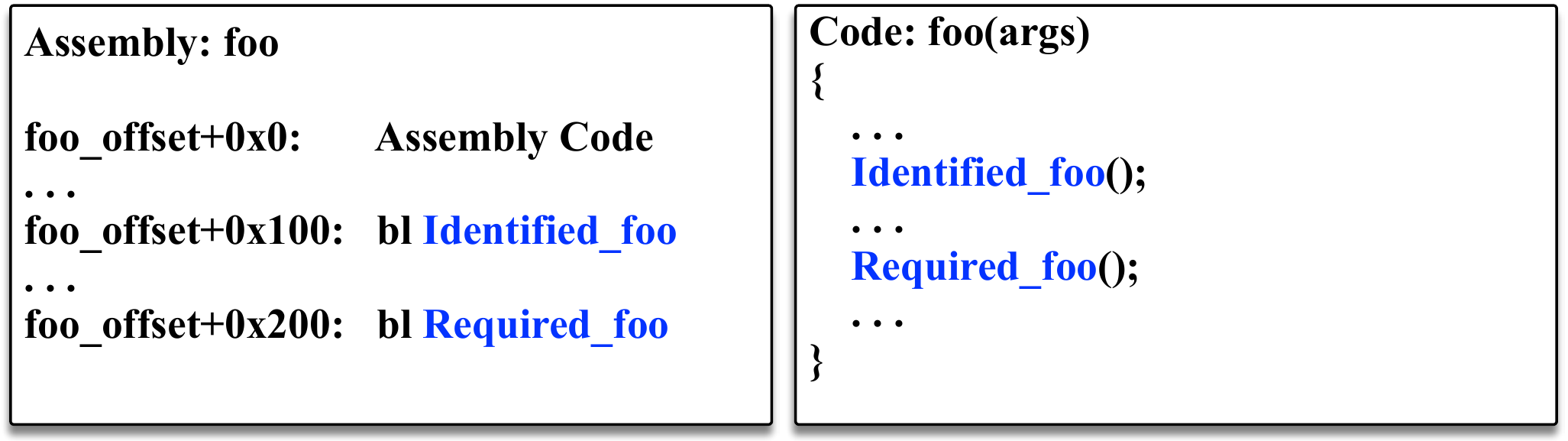}\label{fig:strategy_2_3}}
	\caption{Strategy-II: Function relationship}
	\label{fig:strategy_2}
\end{figure}

\smallskip
\noindent \textbf{Strategy-II: Function relationship}\tab
The second strategy uses the relationship between functions.
That's because functions that do not contain
specific strings cannot be identified by the strategy-I.
However, we can use the relationship between the functions we want
to identify and the ones that have been identified using the previous
strategy. For instance, if we have identified the
function (\code{Identified\_foo}) and this function is \textit{only} invoked  by the
function \code{Required\_foo}, then we can easily locate the \code{Required\_foo}
by finding the caller of the \code{Identified\_foo} function (Figure~\ref{fig:strategy_2_1}).
Similar strategy can be applied to the callee and sibling relationship,
as shown in Figure~\ref{fig:strategy_2_2} and Figure~\ref{fig:strategy_2_3}, respectively. 
Thus, we can identify the functions indirectly with the help of function relationship.



\begin{figure}[t]
	\centering
	
	\subfigure[Logic operation: The constants (i.e., 0x300, -22) of logic operation or return value in source code map to the operands in assembly code.]
	{\includegraphics[width=0.47\textwidth]{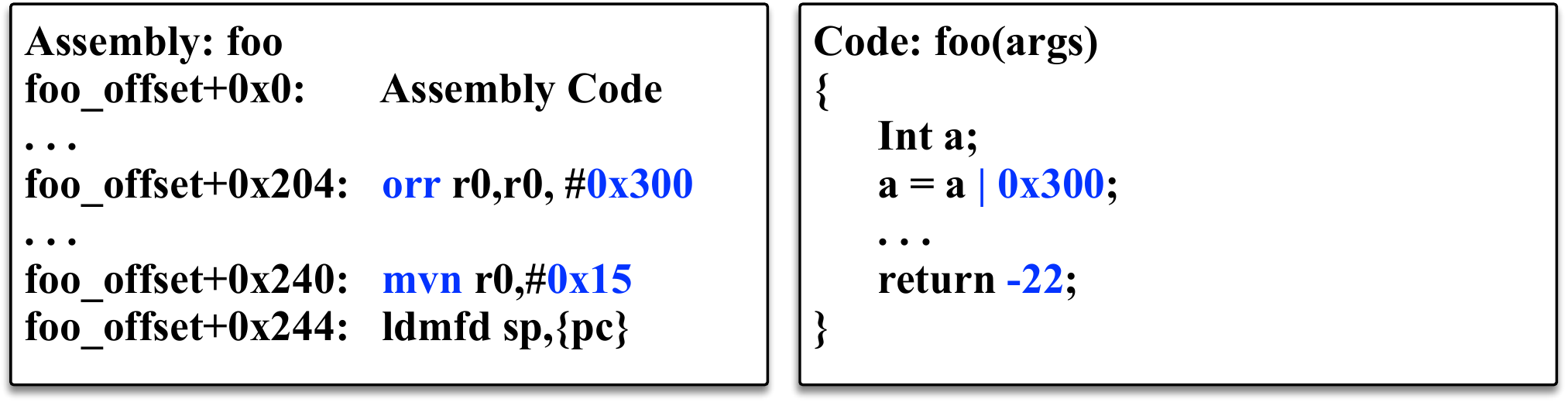}\label{fig:strategy_3_1}}
	
	\subfigure[Callee Number: The two callee functions (i.e., callee\_foo\_one, callee\_foo\_two) map to the two bl instruction at offset foo\_offset+0x18 and foo\_offset+0x1c. Basic Block Number: The three basic blocks in source code maps to three basic blocks in assembly code. ]{\includegraphics[width=0.47\textwidth]{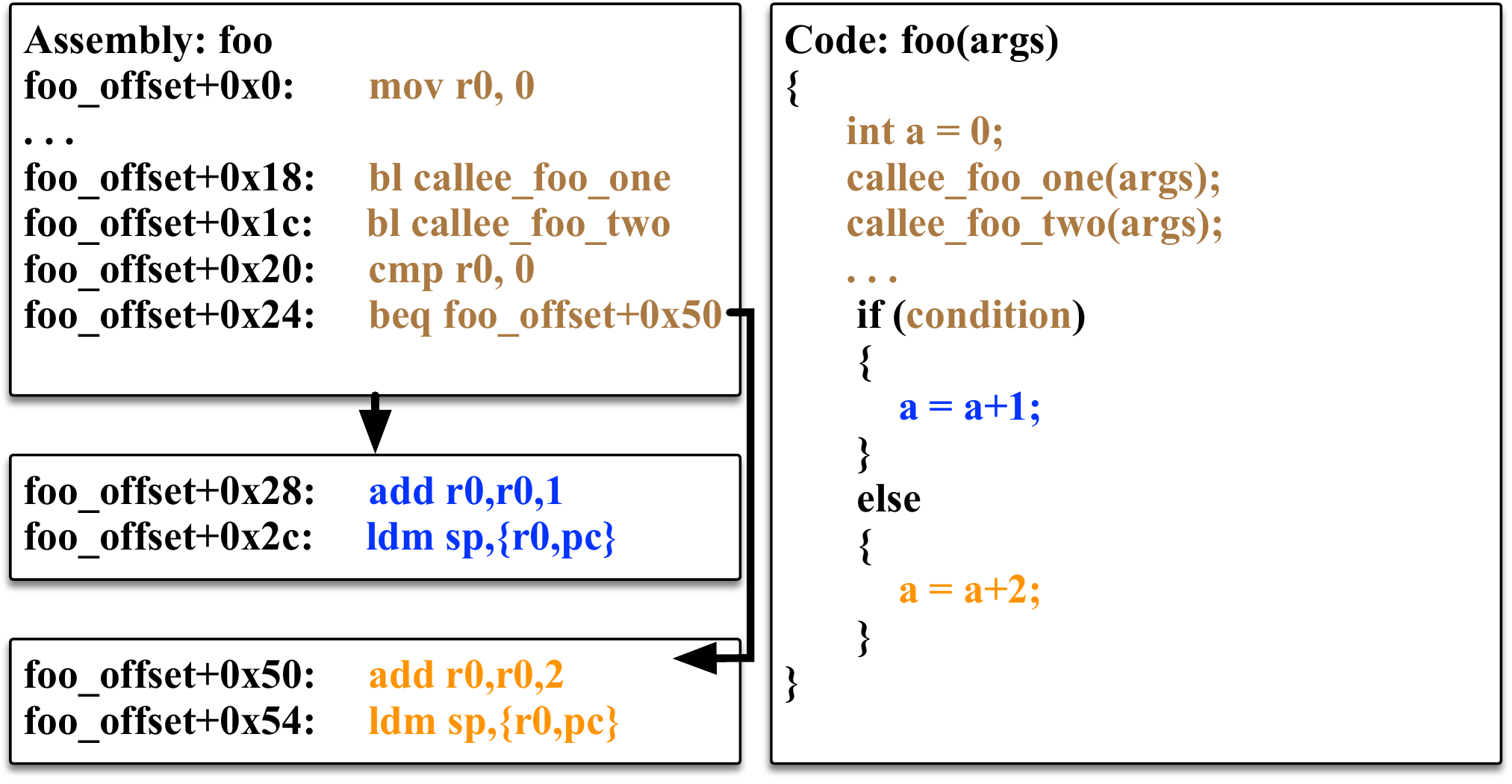}\label{fig:strategy_3_2}}

	\caption{Strategy-III: Function structure}
	\label{fig:strategy_3}
	
\end{figure}

\smallskip
\noindent \textbf{Strategy-III: : Function structure}\tab
If one function has more than one caller, callee or sibling,
it cannot be located solely using the function relationship.
The third strategy takes the function structure, including
logic or arithmetic operations, return value, the number of basic blocks, and
the number of callee functions. 
Fig.~\ref{fig:strategy_3_1} shows the example that the function performs the logic operation on some
specific values (i.e., \code{a = a|0x300}) and return a specific value (i.e., -22) , the compiler will generate
the instructions that contain the specific values (e.g.,
\code{orr r0,r0, \#0x300}, \code{mvn r0,\#0x15}). Besides, the callee number and basic block number will also be considered to filter out the candidate. Fig.~\ref{fig:strategy_3_2} shows that function \textit{foo} has two callees (i.e., \textit{callee\_foo\_one} and \textit{callee\_foo\_two}), which map to two instructions at foo\_offset+0x18 and foo\_offset+0x1c. Basic block number works with the same rule.

\smallskip
\noindent \textbf{Summary}\tab
With the above three strategies, we can automatically and successfully identify ECMO Pointers
for all the Linux kernels ($815$ ones in $20$ kernel versions)
used in the evaluation (Section~\ref{sec:eva_pointers}).

\subsection{Generate ECMO Drivers}
\label{sec:generate_driver}
The process to generate \textit{ECMO Drivers} is similar with developing a kernel
module. However, we need to make the driver self-contained as much as possible
and invoke the APIs in the Linux kernel through \textit{ECMO Backward Pointers}.
In particular, we compile the source code into an object
file (i.e., \code{ECMO\_Driver.o}). To make this driver work, we need to setup the
base address and  fix up
the function calls to \textit{ECMO Backward Pointers}. Moreover, we need to ensure that this driver
does not occupy the physical memory region that the kernel can perceive, which is achieved by allocating the opaque memory.

\smallskip
\noindent \textbf{Fixup the driver}\tab
Note that the compiled object file's base address is \code{0x0}.
Given a new load address at runtime, our system calculates new values of
the data pointers and function pointers and automatically rewrite the corresponding values in
the driver.

Furthermore, due to the limitation of the jump range for the \code{BL Label}
instruction, the driver may not
be able to invoke the functions (\textit{ECMO Backward Pointers}) in the original Linux kernel with direct calls,
if the offset between them is far from the range of the \code{BL} instruction.
To make it work, we rewrite the direct calls with indirect calls.
For example, Fig.~\ref{fig:ecmo_driver} shows a code snippet of the assembly code.
At the offset \code{0x10000}, it loads the value stored at the offset \code{0x10050}
into the register \code{R3}, which is the jump target.
We can rewrite the value in the offset \code{0x10050} to invoke arbitrary
function (\textit{ECMO Backward Pointers}) in the Linux kernel, without being limited by the direct call.

\begin{figure}[t]
	\centering
	\begin{lstlisting}
    0x10000: ldr r3, [pc, #72] 
    0x10004: blx r3
    0x10050: // Pointer value of called function	 
	\end{lstlisting}
\caption{\code{\sysname Driver} indirectly invokes functions in Linux kernel. In offset 0x10000, the memory address pointed by [pc, \#72] is 0x10000 + 8 + 72 = 0x10050. In this case, functions with arbitrary address can be invoked.}
\label{fig:ecmo_driver}
\end{figure}

\begin{figure}[t]
	\centering
	\includegraphics[width=0.6\linewidth]{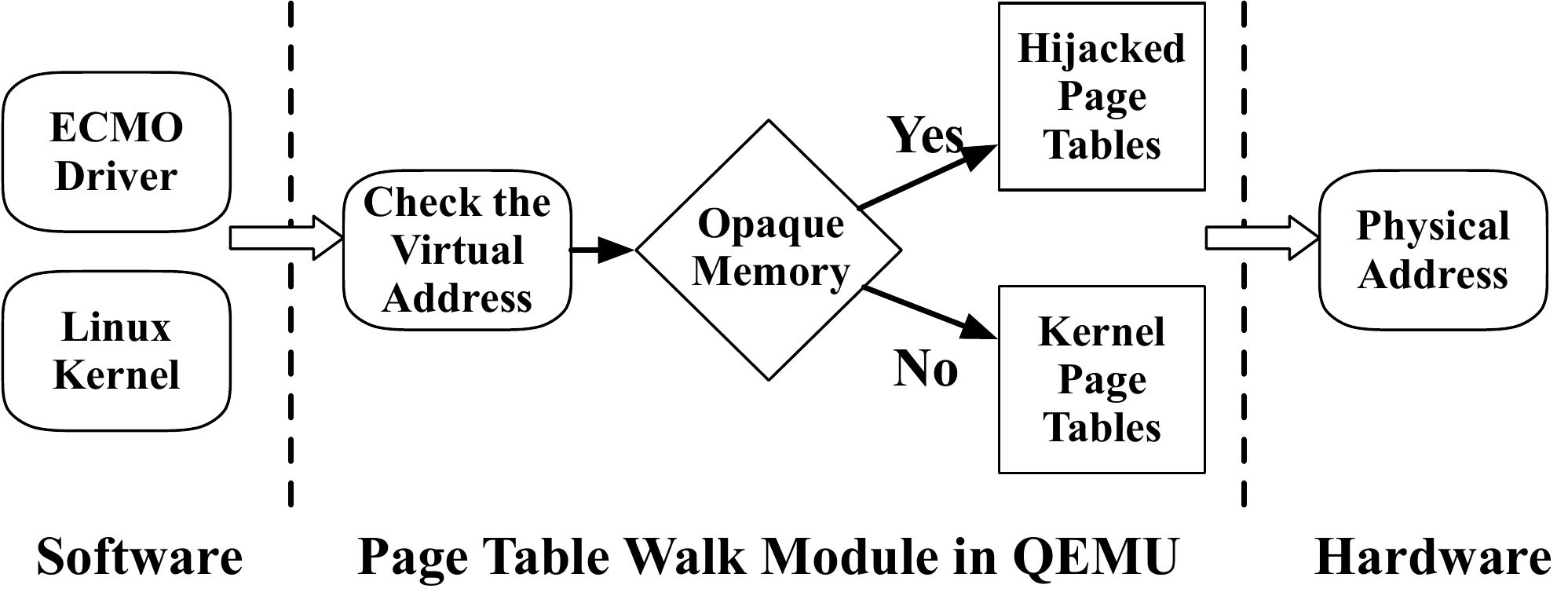}
	
	\caption{The overall design of {opaque memory}.}
	\label{fig:opaque_memory}

\end{figure}

\smallskip
\noindent \textbf{Allocate the opaque memory}\tab
The \textit{ECMO Driver} is loaded into the memory for execution.
As a result, if we directly inject the driver into the physical memory pages
that are treated as free page by the rehosted Linux kernel, the pages could
be allocated for other purposes.
That's because the rehosted kernel does not explicitly know the existence
of the \textit{ECMO Driver} code in the memory.
Thus, we need to guarantee that the driver should reside inside a memory region that
cannot be affected by the Linux kernel.

To solve this problem, we propose the concept of the \textit{opaque memory},
a memory region that is not perceived by the Linux kernel but can be
used at runtime. We implement the {opaque memory} by hooking into
the emulated MMU in QEMU. Fig.~\ref{fig:opaque_memory} shows
how {opaque memory} works.

Specifically, the emulated MMU walks through the page table to translate
virtual addresses to physical addresses. ECMO changes the MMU module
in QEMU to check whether the virtual address being translated is
in the region of the {opaque memory}. If so, it will walk through
our hijacked page table for the {opaque memory}
to get the physical address. Otherwise, the original kernel page tables will be used. 
We ensure that the virtual
address in the {opaque memory} always has a valid entry in the page table. 
By doing so, the \textit{ECMO Driver} can be loaded and executed in the
{opaque memory}, without affecting the memory view of the rehosted Linux kernel.

\subsection{Implementation Details}

We implement \sysname based on LuaQEMU~\cite{luaqemu}. LuaQEMU is a dynamic analysis framework based on QEMU that exposes several QEMU-internal APIs to LuaJIT~\cite{LuaJIT} core, which is injected into QEMU. We port LuaQEMU based on old QEMU (version 2.9.50) to support the QEMU in new version (4.0.0) and exposes more designated APIs for initializing the peripheral models.
With LuaQEMU, we are able to hijack the execution process of 
rehosted Linux kernel at runtime and manipulate the machine states,
e.g., accessing registers and memory regions, through Lua scripts, at specified breakpoints. For example, we can specify a breakpoint at any particular addresses. Inside the breakpoint, we can execute our own Lua script for different purpose. 
This eases the implementation of the {opaque memory}, dump the decompressed Linux kernel, and install the ECMO Pointers.

The module to identify ECMO Pointers (Section~\ref{sec:identify_ecmo_pointers})
is implemented in Python. 
We utilize {Capstone}~\cite{capstone} to disassemble the decompressed
Linux kernel. For the function identification, we re-implement the algorithm described in Nucleus~\cite{andriesse2017compiler} and angr~\cite{angr} with Python. We further extract the required function information, which is the function signature based on the generated functions and their control flow graphs. Finally, we integrate all these code with our strategies for identifying ECMO Pointers, which takes 2290 lines of Python code.
All the above mentioned procedures can be done automatically except that
the \textit{ECMO Driver} is developed using the C language manually, which takes 600 lines of code, and 
cross-compiled by GCC. Note that it is one-time effort to develop the \textit{ECMO Driver}.
One \textit{ECMO Driver} can be used by different Linux kernel versions if the related functions and structures are not changed.


\section{Evaluation}

In this section, we present the evaluation result of our system.
Note that, the main purpose of our work is to rehost Linux kernels
in QEMU so that we can build different dynamic analysis applications and install drivers for more peripherals. In the
following, we first introduce the dataset of firmware images used in
the evaluation and then answer the following
research questions. 

\begin{itemize}[leftmargin=*]
 \setlength{\itemsep}{1pt}
	\item \textbf{RQ1:} Is \sysname able to identify ECMO Pointers?
	\item \textbf{RQ2:} Is \sysname able to rehost the Linux kernels of
	embedded devices with different kernel versions
	and device models?
	\item \textbf{RQ3:} Are the rehosted Linux kernels stable and reliable?
	\item \textbf{RQ4:} Can \sysname support more peripherals and be used to develop dynamic analysis applications?
\end{itemize}

\subsection{Dataset}
As our system targets embedded Linux kernels, we have collected the firmware
images from both third-party projects (i.e., OpenWRT~\cite{openwrt}) and
device vendors (i.e., Netgear~\cite{Netgear}).
Our evaluation targets  Linux kernels in ARM devices, since they
are the popular CPU architectures in embedded devices~\cite{arm_roadshow}.
However, the overall methodology can also be applied to other architectures (e.g., MIPS).

During the experiment, we focuses on transplanting three early-boot peripherals, i.e., interrupt controller (IC),
timer, and UART, which are required to boot a Linux kernel. 
Once the Linux kernel is rehosted, we can install different peripheral drivers  to support other peripherals with kernel modules.
Specifically, we use the PrimeCell Vectored Interrupt Controller (PL190)~\cite{pl190}
and ARM Dual-Timer Module (SP804)~\cite{sp804}.
We use the ns16550 UART device in our system.
In total, we evaluate $815$ ($720$ in OpenWRT and $95$ in Netgear) firmware images that contain Linux kernels.

\subsection{Identify ECMO Pointers (RQ1)}
\label{sec:eva_pointers}
\begin{table}
\centering
\footnotesize
\caption{The \sysname Pointers, identification strategy, and the Linux kernel versions that the  \sysname pointers used by. }

\begin{tabular}{lcc} 
\toprule
Forward Pointers                   & Strategy & Kernel Version       \\ 
\hline
init\_irq               & I        & ALL                  \\ 

init\_time              & I        & ALL                  \\ 

\hline
Backward Pointers                  &     Strategy       &       Kernel Version           \\ \hline

irq\_set\_chip\_and\_handler\_name & III      & 3.18.x/4.4.x/4.14.x  \\ 

irq\_set\_chip\_data               & III      & ALL                  \\ 

handle\_level\_irq                 & II       & ALL                  \\ 

\_\_handle\_domain\_irq            & III      & 3.18.x/4.4.x/4.14.x  \\ 

setup\_machine\_fdt                & I        & 3.18.x/4.4.x/4.14.x  \\ 

set\_handle\_irq                   & III      & 3.18.x/4.4.x/4.14.x  \\ 

irq\_domain\_add\_simple           & III      & 3.18.x/4.4.x/4.14.x  \\ 

irq\_create\_mapping               & I        & 3.18.x/4.4.x/4.14.x  \\ 

of\_find\_node\_by\_path           & II       & 3.18.x/4.4.x/4.14.x  \\ 

setup\_irq                         & I        & ALL                  \\ 

clockevents\_config\_and\_register & III      & 3.18.x/4.4.x/4.14.x  \\ 

irq\_domain\_xlate\_onetwocell     & I        & 3.18.x/4.4.x/4.14.x  \\ 

clockevent\_delta2ns               & I        & 2.6.x                \\ 

clockevents\_register\_device      & II       & 2.6.x                \\ 

set\_irq\_flags                    & I        & 2.6.x/3.18.x         \\ 

set\_irq\_chip                     & I        & 2.6.x                \\ 

irq\_to\_desc                      & II       & 2.6.x                \\ 

\_\_do\_div64                      & II       & 2.6.x                \\ 

platform\_device\_register         & I        & ALL                  \\ 

lookup\_machine\_type              & I        & 2.6.x                \\ 

\_set\_irq\_handler                & I        & 2.6.x                \\ 

irq\_modify\_status                & III      & 4.4.x/4.14.x         \\
\bottomrule
\end{tabular}

\label{tab:pointer}

\end{table}
ECMO Pointers are important to {peripheral transplantation}.
In this section, we evaluate the success rate of identifying {\sysname Pointers}. 
Among all the $815$ Linux kernels, there are $20$ different kernel versions.

Table~\ref{tab:pointer} lists the required ECMO Pointers,
the strategies we used, and the Linux kernel versions that these ECMO
Pointers are used. In total, we need to identify $24$ different ECMO Pointers
for all the $20$ Linux kernel versions. Among them,
two (i.e., \code{mach\_desc-\textgreater{}init\_time}, and \code{mach\_desc-\textgreater{}init\_irq} ) are data pointers.
Identifying the data pointers is rather more difficult than the function pointers
as we need to identify symbols in each function and infer the right ones.
Fortunately, these two data pointers are the return values of \code{setup\_machine\_fdt} and \code{lookup\_machine\_type}, respectively. According to the ARM calling convention, the return value is saved in register \code{R0}. In this case, we can identify these two data pointers by identifying function pointers \code{setup\_machine\_fdt} and \code{lookup\_machine\_type}.

\begin{table}[t]
\centering
\caption{The decompressed Linux kernel size and the disassembled function numbers for our dataset.}

\begin{tabular}{ccccc} 
\toprule
                                  & Maximum  & Minimum & Mean    & Median   \\ 
\hline\hline

Size (Bytes)               & 8,526,240  & 4,134,392 & 7,297,977 & 8,478,848  \\ 

Functions (\#) & 48,412    & 18,455   & 29,910   & 23,872    \\
\bottomrule
\end{tabular}

\label{tab:disa}
\end{table}
Identifying ECMO Pointers requires us to disassemble the decompressed Linux kernel.
Table~\ref{tab:disa} lists the information of these kernels.
The decompressed Linux kernel is about 730k bytes on average, with thousands of functions.
Among these functions, we successfully identify the required ECMO Pointers for all
Linux kernels.

\begin{table*}[t]
\centering
\footnotesize
\setlength{\extrarowheight}{0pt}
\addtolength{\extrarowheight}{\aboverulesep}
\addtolength{\extrarowheight}{\belowrulesep}
\setlength{\aboverulesep}{0pt}
\setlength{\belowrulesep}{0pt}
\caption{The overall result of \sysname on rehosting the Linux kernel of OpenWRT. "Downloaded Images" represents the number of downloaded images. "Format Supported" represents the number of images whose formats are supported by firmware extraction tool (i.e., Binwalk). "Kernel Extracted" represents the number of images extracted from the downloaded image, which are rehosted by ECMO. "Peripherals Transplanted" represents the number of the images that peripheral can be transplanted successfully (e.g., IC can handler the interrupt well). "Ramfs are not Mounted" represents the number of images that cannot mount the given ramfs.
"Shell" represents the images that we can rehost and spawn a shell. 
Success Rate of Transplantation = (Peripherals Transplanted)/(Images); Success Rate of Rehosting = (Shell)/(Images).}

\scalebox{0.78}{
\begin{tabular}{ccccccccc} 
\toprule
Kernel  Version & Downloaded Images & Format Supported & Kernel Extracted &\begin{tabular}[c]{@{}c@{}}Peripherals \\ Transplanted \end{tabular} & \begin{tabular}[c]{@{}c@{}}Success Rate of\\ Transplantation \end{tabular} & \begin{tabular}[c]{@{}c@{}}Ramfs are \\ not Mounted \end{tabular}& Shell & \begin{tabular}[c]{@{}c@{}}Success Rate of \\ Rehosting \end{tabular}  \\ 
\hline\hline
3.18.20  &    23&23   & 21     & 21                                                                   & {\cellcolor[rgb]{0.643,0.863,1}}100\%       &8                               & 13    & {\cellcolor[rgb]{0.643,0.863,1}}61.9\%                                   \\ 

3.18.23  &    29&29   & 29     & 29                                                                   & {\cellcolor[rgb]{0.643,0.863,1}}100\%       &8                               & 21    & {\cellcolor[rgb]{0.643,0.863,1}}72.4\%        
\\ 

4.4.42  &    37&37   & 37     & 37                                                                  & {\cellcolor[rgb]{0.643,0.863,1}}100\%       &8                               & 29   & {\cellcolor[rgb]{0.643,0.863,1}}78.4\%                                   \\ 

4.4.47  &    37&37   & 37    & 37                                                                  & {\cellcolor[rgb]{0.643,0.863,1}}100\%       &8                               & 29    & {\cellcolor[rgb]{0.643,0.863,1}}78.4\%                                   \\ 

4.4.50     & 45 & 45     & 45   & 45                                                                  & {\cellcolor[rgb]{0.643,0.863,1}}100\%                                &16      & 29    & {\cellcolor[rgb]{0.643,0.863,1}}64.4\%                                 \\ 

4.4.61     & 39 & 39     & 37     & 37                                                                   & {\cellcolor[rgb]{0.643,0.863,1}}100\%                            & 8          & 29    & {\cellcolor[rgb]{0.643,0.863,1}}78.4\%                                 \\ 

4.4.71    & 40 & 40      & 38     & 38                                                                   & {\cellcolor[rgb]{0.643,0.863,1}}100\%                                  &8    & 30    & {\cellcolor[rgb]{0.643,0.863,1}}78.9\%                                 \\ 

4.4.89  &40 &40        & 38     & 38                                                                   & {\cellcolor[rgb]{0.643,0.863,1}}100\%                             &8         & 30    & {\cellcolor[rgb]{0.643,0.863,1}}78.9\%                                 \\ 

4.4.92    &41 & 41      & 38     & 38                                                                   & {\cellcolor[rgb]{0.643,0.863,1}}100\%                            &8          & 30    & {\cellcolor[rgb]{0.643,0.863,1}}78.9\%                                 \\ 

4.4.140   &41 &41      & 38     & 38                                                                   & {\cellcolor[rgb]{0.643,0.863,1}}100\%                             &8         & 30    & {\cellcolor[rgb]{0.643,0.863,1}}78.9\%                                 \\ 

4.4.153  &40 &38       & 38     & 38                                                                   & {\cellcolor[rgb]{0.643,0.863,1}}100\%                        &8              & 30    & {\cellcolor[rgb]{0.643,0.863,1}}78.9\%                                 \\ 

4.4.182   &40&38      & 38     & 38                                                                   & {\cellcolor[rgb]{0.643,0.863,1}}100\%                            &8          & 30    & {\cellcolor[rgb]{0.643,0.863,1}}78.9\%                                 \\ 

4.14.54   &54 &54      & 42     & 42                                                                   & {\cellcolor[rgb]{0.643,0.863,1}}100\%                             &0         & 42    & {\cellcolor[rgb]{0.643,0.863,1}}100\%                                  \\ 

4.14.63     &66 &66   & 42     & 42                                                                   & {\cellcolor[rgb]{0.643,0.863,1}}100\%                          &0            & 42    & {\cellcolor[rgb]{0.643,0.863,1}}100\%                                  \\ 

4.14.95    &66&66     & 42     & 42                                                                   & {\cellcolor[rgb]{0.643,0.863,1}}100\%                            &0           & 42    & {\cellcolor[rgb]{0.643,0.863,1}}100\%                                  \\ 

4.14.128   &66&66     & 42     & 42                                                                   & {\cellcolor[rgb]{0.643,0.863,1}}100\%                            &0           & 42    & {\cellcolor[rgb]{0.643,0.863,1}}100\%                                  \\ 

4.14.131   &66&66     & 42     & 42                                                                   & {\cellcolor[rgb]{0.643,0.863,1}}100\%                            &0           & 42    & {\cellcolor[rgb]{0.643,0.863,1}}100\%                                  \\ 

4.14.151    &66&66    & 42     & 42                                                                   & {\cellcolor[rgb]{0.643,0.863,1}}100\%                            &0           & 42    & {\cellcolor[rgb]{0.643,0.863,1}}100\%                                  \\ 

4.14.162  &66&66      & 42     & 42                                                                   & {\cellcolor[rgb]{0.643,0.863,1}}100\%                            &0           & 42    & {\cellcolor[rgb]{0.643,0.863,1}}100\%                                  \\ 
\hline
Overall  &902 &898       & 720    & 720                                                                  & {\cellcolor[rgb]{0.643,0.863,1}}100\%                               &96       & 624   & {\cellcolor[rgb]{0.643,0.863,1}}86.7\%                                 \\
\bottomrule
\end{tabular}}

\label{tab:overall}
\end{table*}

\begin{tcolorbox}[size=title]
{\textbf{Answer to RQ1: } ECMO can identify all the required ECMO Pointers from thousands of functions inside decompressed Linux kernel.}
\end{tcolorbox}

\subsection{Rehost Linux Kernels (RQ2)}
In this section, we evaluate the capabilities of \sysname on rehosting
the Linux kernels. During this process, we use our system to boot the kernel
and provide a root file system (rootfs) in the format of ramfs. We use our own rootfs because we can include different benchmark applications into the rootfs to conduct security analysis. For example, we include PoCs of kernel exploits to conduct the root cause analysis (Section~\ref{sec:application}). Furthermore, we can include different peripheral drivers to support more peripherals.
The rootfs extracted from the firmware image can also be used.

\subsubsection{Firmware Images from Third Party Projects}

Table~\ref{tab:overall} shows the overall result and the success
rate of {peripheral transplantation} and kernel rehosting for OpenWRT.
We define the success of  {peripheral transplantation} as that
the transplanted IC, timer and UART devices function well in the kernel.
If the rehosted kernel enters into the user-space and spawns a shell,
we treat it as a successful kernel rehosting. In total, we download 902 firmware images from OpenWRT. However, four images' formats are not supported by Binwalk and the Linux kernel cannot be extracted (if there is). For the left 898 firmware images, 720 of them contain Linux kernels while the left ones contain only user-level applications.  The 720 ones will be evaluated by \sysname.

\smallskip
\noindent \textbf{Linux Kernel Versions}\tab
The kernels in the $720$ OpenWRT firm\-ware images consist of 
$19$ different kernel versions. Our evaluation shows that we can
transplant the peripherals  for all the 720 Linux kernels. 
However, some Linux kernels cannot be booted. This is because they
cannot recognize our pre-built root file system (in the ramfs file format) as the support of ramfs is not enabled when being built.
Without the root file system, we cannot launch the shell. However, all of them  enter into the function (i.e., \code{init\_post}) to
execute the \code{init} program.
In summary, among $720$ kernels, our system can rehost $624$ of them, which is shown in Table~\ref{tab:overall}. 

\begin{figure}[t]
	\centering
	\includegraphics[scale=0.65]{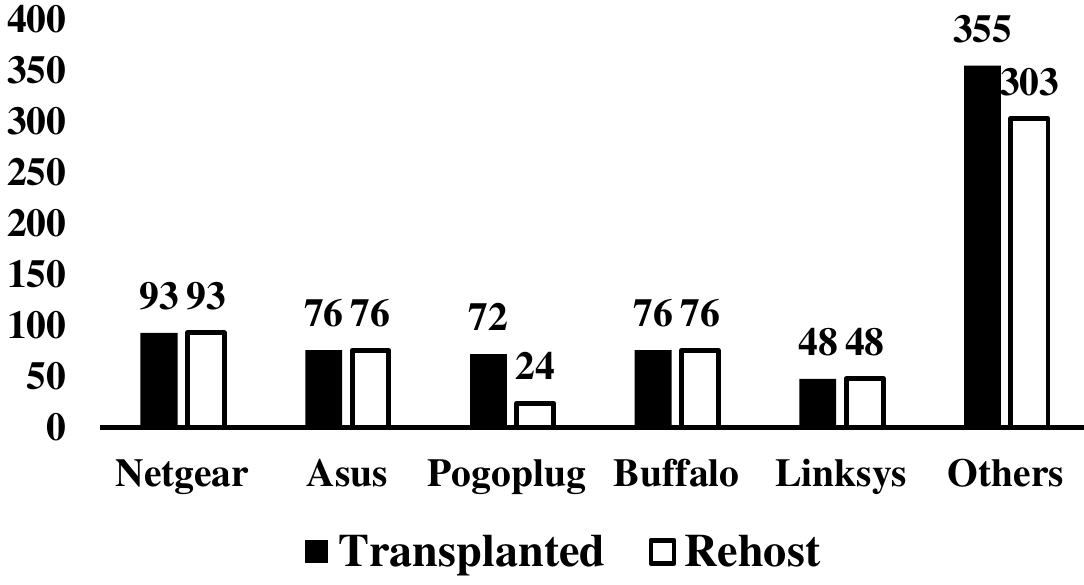}
	
	\caption{Supported Vendors of OpenWRT  Linux Kernels.}
	\label{fig:vendor}
\end{figure}

\smallskip
\noindent \textbf{Vendors and Device Models}\tab
As the OpenWRT project supports devices from multiple vendors, we
calculate the supported vendors and there are 24 different vendors.
Figure~\ref{fig:vendor} shows the result of the top five vendors,
i.e., \code{Netgear}, \code{Asus}, \code{Pogoplug}, \code{Buffalo}, and \code{Linksys},
in the OpenWRT dataset. Among them, \code{Pogoplug} has a relatively
low success rate of rehosting. That's because most kernels from that vendor
cannot recognize our pre-built root file system. We also count the number of device models for the successfully rehosted Linux kernels. In total, 32 device models are identified.

\begin{table}
\centering
\footnotesize
\caption{The overall result of \sysname on rehosting the Linux kernel of Netgear Devices.}

\begin{tabular}{ccccc} 
\toprule
Device Name & Kernel Version & Images & \# of Peripherals Transplanted & Shell  \\ 
\hline
\hline
R6250       & 2.6.36         & 21     & 21                        & 15     \\ 

R6300v2     & 2.6.36         & 22     & 22                        & 19     \\ 

R6400       & 2.6.36         & 20     & 20                        & 20     \\ 

R6700       & 2.6.36         & 16     & 16                        & 16     \\ 

R6900       & 2.6.36         & 16     & 16                        & 16     \\ \hline
Overall & - & 95 & 95 & 86 \\
\bottomrule
\end{tabular}

\label{tab:netgear}

\end{table}
\subsubsection{Firmware Images from Official Vendors}
Besides third-party firmware images, we also apply ECMO on the
official images released by \code{Netgear}. We collect the firmware
images for five popular devices, including R6250, R6300v2, R6400, R6700, R6900,
from the vendor's website~\cite{Netgear}. In total, we manage to collect
95 firmware images, and the latest one is released on 2020-09-30.
Table~\ref{tab:netgear} shows
the result. We noticed that all the Linux kernels of these devices are in the version
2.6.36.
We can successfully transplant the peripherals to all the $95$ different firmware
images. Among them, we can launch the shell for $86$ images while the left 9
cannot be rehosted due to the same root file system problem.

\begin{figure*}[t]
	\centering
	\includegraphics[scale=0.28]{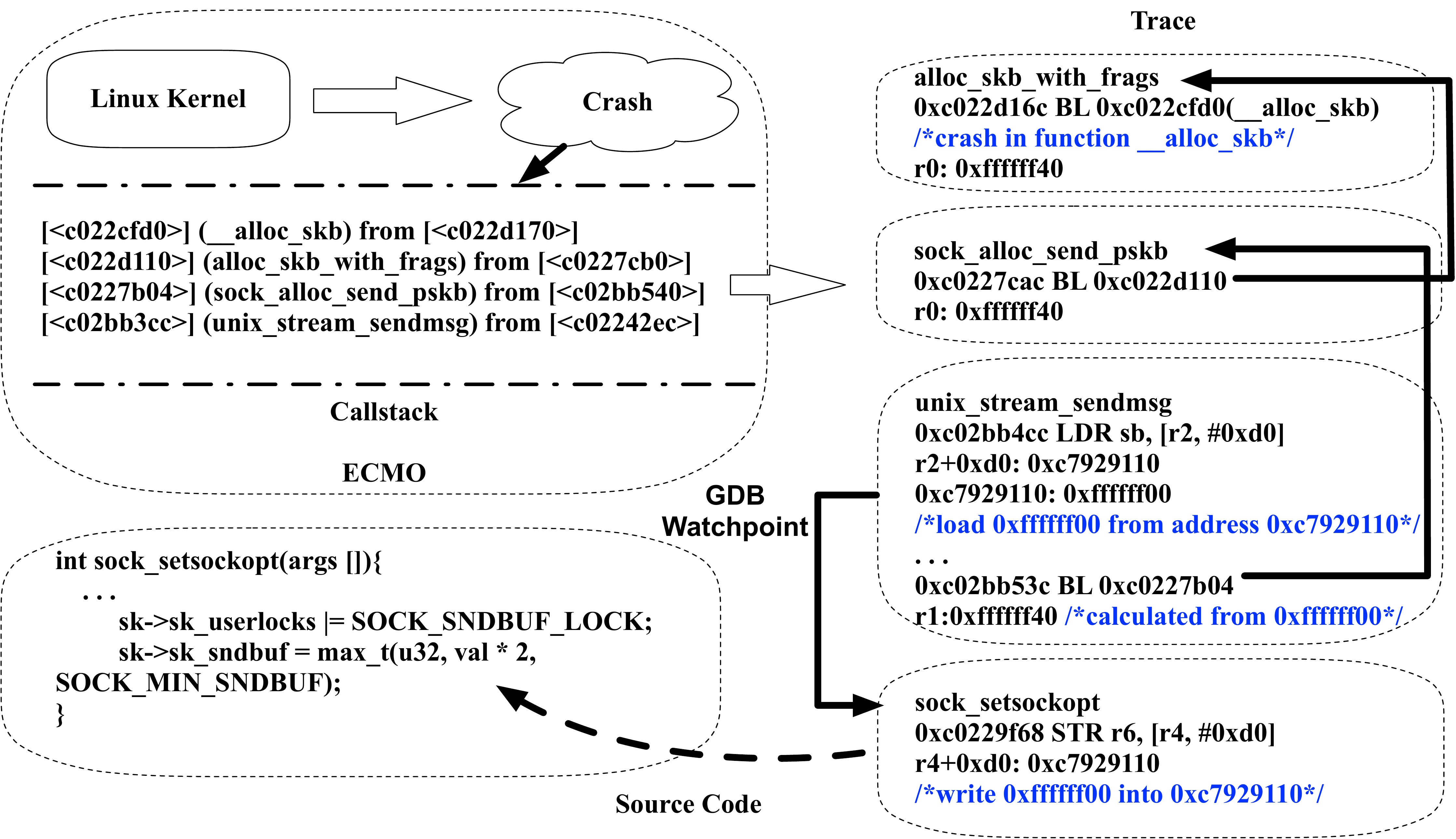}

	\caption{Root cause analysis of CVE-2016-9793.}
	\label{fig:root_cause}
		
\end{figure*}

\begin{tcolorbox}[size=title]
{\textbf{Answer to RQ2: } ECMO can rehost the Linux kernel of embedded devices from 20 kernel versions and 37 (32 in OpenWRT and 5 in Netgear) device models. Peripherals can be transplanted to all the Linux kernels while 87.1\% (710/815) Linux kernels can be successfully rehosted (i.e., launch the shell).}
\end{tcolorbox}

\subsection{Reliability and Stability (RQ3)}
\label{sec:functionality}
\begin{table}[t]
\footnotesize
\centering
\caption{The category of the failed syscall test cases.}

\begin{tabular}{cc} 
\toprule
Category of Failed cases                             & Number  \\ 
\hline
\hline
Testing the bug or vulnerability of Linux kernel & 16      \\ 

Network is not enabled                           & 15      \\ 

The function is not implemented                  & 25      \\ 

Others                                           & 10      \\
\hline
Total                                             & 66   \\
\bottomrule
\end{tabular}
\label{tab:fail_case}

\end{table}
We use the LTP (Linux Test Project~\cite{ltp}) testsuite
to evaluate the reliability and stability of the rehosted kernel.
In total, there are $1,257$ test cases for system calls.
Among them, $148$ are skipped as the testing environment
(e.g., the CPU architecture and the build configuration)
does not meet the requirement.
For the left $1,109$ test cases, $1,043$ 
passed while the left $66$  ones failed. 

We further analyze the reason for the failed test cases.
Table~\ref{tab:fail_case} lists the category of the reason.
Among them, $15$ cases are due to the lack of network devices.
This is expected since our system does not add the support of network device initially. However, all the $15$ test cases are passed
after installing the Ethernet device driver with kernel modules on the rehosted Linux kernel (Section~\ref{subsec:other_peripherals}).
Also, $16$ cases aim to test whether the Linux kernel fixes a bug or vulnerability. 
For instance, the test case (timer\_create03~\cite{timer_create03}) is to check whether 
CVE-2017-18344~\cite{CVE-2017-18344} is fixed. If the vulnerability is
not fixed, the test case will fail. They are also expected since the testing kernel
does not fix these vulnerabilities.
The other $25$ cases return back the \code{ENOSYS} error number, which means
the functionalities are not implemented. For the remaining $10$ cases,
the reason is adhoc, such as the kernel version is old and timeout. 

In summary, $94\%$ of the system call test cases passed.  This evaluation shows the rehosted kernel is reliable and stable.
We further demonstrate the usage scenarios of the rehosted Linux kernel in Section~\ref{sec:application}.

\begin{tcolorbox}[size=title]
{\textbf{Answer to RQ3: } The rehosted Linux kernel can pass 94\% system call test cases in LTP, which demonstrates its reliability and stability. }
\end{tcolorbox}

\subsection{Applications and Other Peripherals (RQ4)}
\label{sec:application}
Our system can rehost Linux kernels, which provides the capability to install different peripheral drivers with kernel modules to support more peripherals.  Furthermore, the rehosted Linux kernel lays the foundation of applications relying on the capability to
introspect the runtime states of the target system. 
In this section, 
we  successfully install the Ethernet device driver (i.e., smc91x) for all the rehosted Linux kernels. 
We also leverage our system to build three applications,
including kernel crash analysis, rootkit forensic analysis, and
kernel fuzzing, to demonstrate the usage scenarios of ECMO.
Other applications that rely on QEMU can be
ported. 
Note that, we only use these applications to demonstrate the usage 
of our system. The applications are not the main contribution of
this work.

\subsubsection{Other Peripherals}
\label{subsec:other_peripherals}
Linux kernel module is an object file that can be loaded during the runtime to extend the functionality of the Linux kernel. In this case, peripheral drivers can be built as kernel modules and loaded into the kernel dynamically. To demonstrate that our rehosted Linux kernel is able to support more peripherals. we select one rather complex peripheral (i.e., smc91x~\cite{smc91x_src}) and build the driver code into kernel module (i.e., smc91x.ko). We then inject this kernel module into the ramfs that is fed to rehosted Linux kernel. After the embedded Linux kernel is rehosted by \sysname, we use the command \textit{insmod smc91x.ko} to install the peripheral driver for smc91x. Meanwhile, QEMU has already provided the peripheral model for smc91x and we can integrate this model into the machine model directly. Finally, we successfully install the peripheral driver of smc91x for all the 710 rehosted Linux kernels, which demonstrate the capability of \sysname to support the other peripherals.

\begin{table}[t]
\footnotesize
\centering
\caption{CVEs that can be triggered on the rehosted Linux kernel by \sysname.}

\begin{tabular}{cccc}
\toprule
CVE ID & CVE Score & CVE Type & Fix Version \\ \hline

CVE-2018-5333  & 5.5 & Null Pointer Dereference & 4.14.13 \\ 
CVE-2016-4557  & 7.8 & Double Free              & 4.5.5   \\ 
CVE-2017-10661 & 7.0 & Race Condition           & 4.10.15 \\ 
CVE-2016-0728  & 7.8 & Interger Overflow        & 4.4.1   \\ 
CVE-2016-9793  & 7.8 & Type Confusion           & 4.8.14  \\ 
CVE-2017-12193 & 5.5 & Null Pointer Dereference & 4.13.11 \\ \bottomrule
\end{tabular}

\label{tab:cve}

\end{table}
\subsubsection{Crash Analysis}
In the following, we show the process to utilize \sysname to  
understand the root cause of the crash on rehosted kernels. 
To this end, we collect the PoCs that can trigger the crash for six
reported bugs and vulnerabilities (as shown in Table~\ref{tab:cve}).
We then boot the Linux kernel and
run the PoCs to crash the kernel. During this process, we use the QEMU
to collect the runtime trace. We also leverage the remote GDB in QEMU
to debug the rehosted kernel. 
We detail the procedures on how to conduct the crash analysis for one case (CVE-2016-9793~\cite{cve_2016-9793}) with the collected runtime trace.
Figure~\ref{fig:root_cause} shows the whole procedure.

Specifically, when the rehosted Linux kernel crashes, the detailed call
stack will be printed out. The call stack includes the function name
and the addresses of these functions. 
With the runtime trace provided by QEMU,
we can get the information including the
register values and the execution path.
By analyzing the trace, we noticed that a negative value
(i.e., \code{0xffffff40}) is the first parameter of the function
\code{\_\_alloc\_skb}. This negative value results in the crash. 

We then analyze the propagation of this negative value within the trace.
This value is propagated by the first parameter of the function
\code{sock\_alloc\_send\_pskb}. Finally, we notice that the negative
value \code{0xffffff40} is calculated from \code{0xffffff00}, which is loaded by
the function \code{unix\_stream\_sendmsg} from the address \code{0xc7929110}.
We then use the \code{GDB} to set a watchpoint at this memory address
and capture that the instruction at the address \code{0xc0229f68} was writing
the negative value (i.e., \code{0xffffff00} ) into this memory location.

We further analyze the function that contains the instruction at the
address \code{0xc0229f68}. It turns out that the root cause of the crash
is because of the type confusion. In the function \code{sock\_setsockopt},
the variable \code{sk$\rightarrow$sk\_sndbuf} will be set by the return value
of \code{max\_t} (maximum value between two values in the same type).
However, due to the wrong type \code{u32}, the return value can be a negative
value, which triggers the crash.

This analysis shows the usage of ECMO by providing the capability
introspect the runtime states of the rehosted kernel.

\subsubsection{Rootkit Forensic Analysis}
Rootkit forensic analysis requires the ability to monitor the runtime
states of the kernel~\cite{garfinkel2003virtual,jiang2010stealthy}.
We demonstrate this ability by conducting the rootkit forensic analysis
with one (i.e., \textit{Suterusu}~\cite{suterusu}) popular rootkit in the wild. 

\begin{figure}[t]
	\centering
	
	\subfigure[Workflow of rootkit \textit{Suterusu}]{\includegraphics[width=0.37\textwidth]{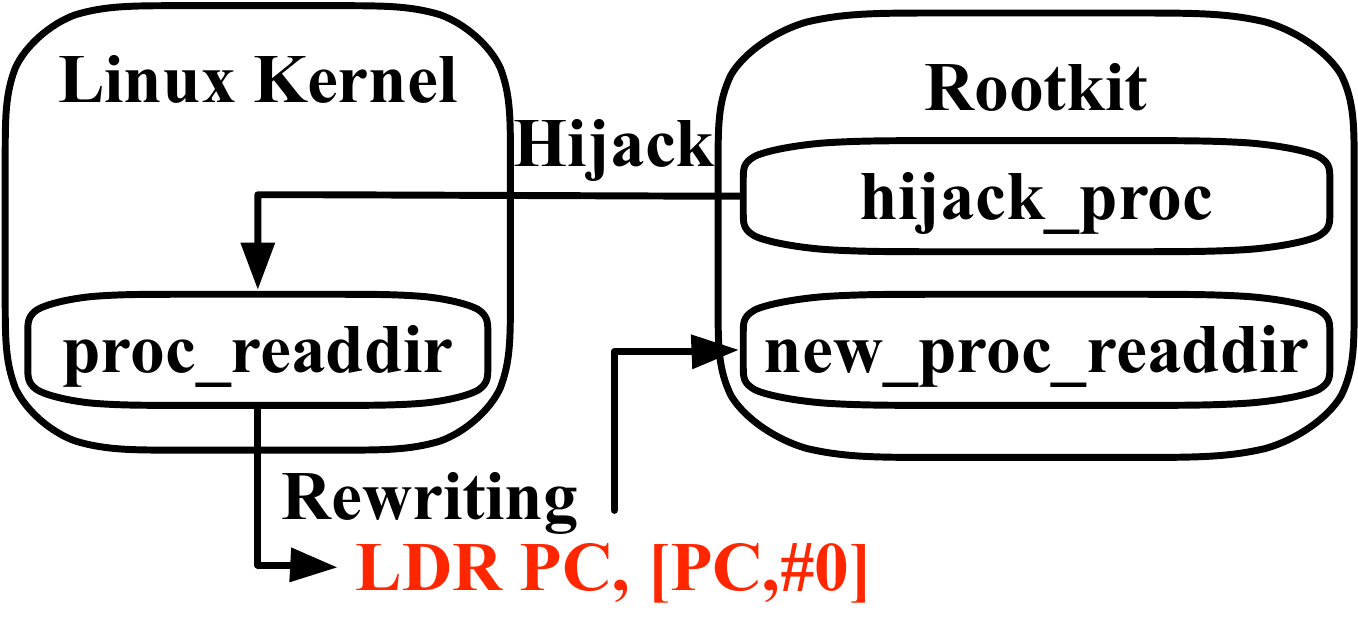}\label{fig:rootkit_flow}}
	\subfigure[\sysname observes how the rootkit \textit{Suterusu}  works. ]{\includegraphics[width=0.3\textwidth]{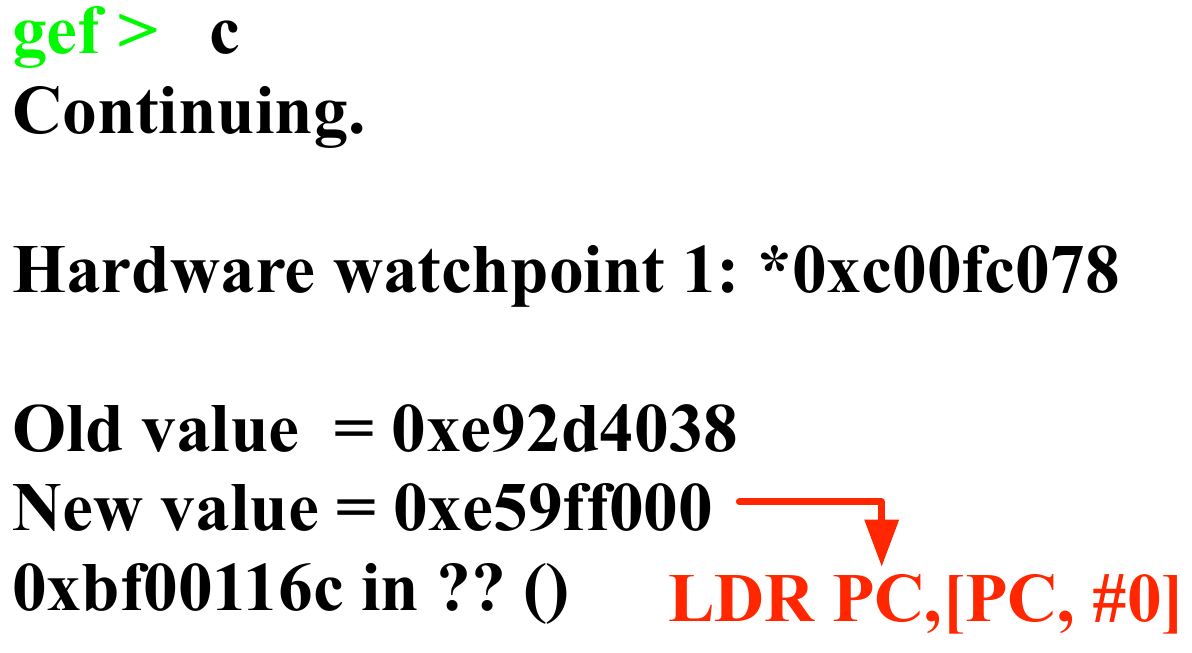}\label{fig:rootkit_gdb}}

	\caption{The workflow of rootkit  \textit{Suterusu} and how \sysname analyzes the behavior}

	\label{fig:rootkit}

\end{figure}

Specifically, \code{Suterusu} is able to hide specific processes by
hijacking the kernel function \code{proc\_readdir}, which is used to get
the process information.
As shown in Figure~\ref{fig:rootkit_flow}, it hijacks the function
\code{proc\_readdir} by rewriting the function's first instruction
to \code{LDR PC,[PC,\#0]}. As a result, it redirects the execution to
the function \textit{new\_proc\_readdir} inside the rootkit.
With ECMO, we can monitor the changes to the kernel code sections
(a suspicious behavior) by setting up memory watchpoints to the
Linux code section (Figure~\ref{fig:rootkit_gdb}).

\subsubsection{Fuzzing}
Fuzzing has been widely used to detect software vulnerabilities.
We ported one of the most popular kernel fuzzers (i.e.,\code{UnicornFuzz}~\cite{maier2019unicorefuzz}) into ECMO and
fuzzed the example kernel modules provided by \code{UnicornFuzz}. 
 \code{UnicornFuzz} can work under ECMO 
and the fuzzing speed can reach to $396$ instances
per second. This demonstrates the usage of ECMO for kernel fuzzing.

\begin{tcolorbox}[size=title]
{\textbf{Answer to RQ4: } Applications, e.g., crash analysis, forensic analysis, kernel fuzzing, can be built upon the rehosted Linux kernel by our system. Furthermore, rehosted Linux kernel can install peripheral drivers with kernel modules to support more peripherals. }
\end{tcolorbox}

  \section{Discussion}
\label{sec:discussion}
\noindent \textbf{Manual efforts}\tab
\sysname provides mostly automated approach and only developing the \textit{ECMO Driver} requires manual efforts. However, this is a one-time effort.
Furthermore, one \textit{\sysname Driver} can be transplanted to different kernel versions if the related functions and structures are not changed. 
Even if the functions are changed, we just need to change a few APIs and compile it again to create a new \textit{\sysname Driver}.
For example, the 815 Linux kernels consist of 20 different kernel versions. For the kernel in version 2.6.36, it takes 385 lines of C code. This driver can be used for all the kernel images of Netgear (Table 4). For the kernel in version 3.18.20 and 3.18.23, it takes 534 lines of C code while 180 lines of new code are added. For kernels in all the left 17 versions, they share the same driver code. 60 lines of new code are added compared with the one used in 3.18.20. Note that  the driver code for the transplanted peripherals does not need to be developed. Instead, we reuse the existing code. For example, the driver code for VIC (PL190) is open source~\cite{pl190_src}. Thus, we just reuse the existing driver code, merge the driver code into one file, and finally compile it to generate the ECMO driver. In total, it takes less than one person-hour to build a new customized driver.


\smallskip
\noindent \textbf{Functionalities of peripherals}\tab
We successfully boot the Linux kernel by transplanting designated
peripherals (e.g., IC, Timer, and UART).
We admit that the original peripherals may not work property as 
they are not emulated (or transplanted) in QEMU. However, the functionalities of the transplanted peripherals are guaranteed. With the transplanted peripherals, \sysname can provide the capability to introspect the runtime states
of the Linux kernel that dynamic analysis applications can be built upon.
Without our system, it's impossible to build such applications since the target
Linux kernel cannot be booted in QEMU.
The three applications used in the evaluation have demonstrated the usage scenarios
of our system. We may build or port more complicated applications, e.g.,  
dynamic taint analysis~\cite{schwartz2010all}, to further evaluate our system.


\smallskip
\noindent \textbf{Other peripherals}\tab
Currently, \sysname is evaluated based on transplanting three early-boot peripherals (i.e., IC, timer, and UART) as they are required to boot a Linux kernel. In general, peripheral transplantation works on all kinds of peripherals. The transplanting process depends on the identification of ECMO pointers. Fortunately, to support the other peripherals, users can install the kernel modules directly on the rehosted Linux kernel, which does not need to identify pointers.
In this case, all kinds of peripherals can be supported. Our experiments show that the driver of Ethernet device, which is rather complex, can be successfully installed and the network functionality can be guaranteed.


\smallskip
\noindent \textbf{Other architectures}\tab Currently, \sysname only supports ARM architecture, which is the most popular one in embedded systems~\cite{arm_roadshow}. However, the technique peripheral transplantation can be easily extended to the other architecture as it does not rely on any particular architecture feature. Specifically, developers need to implement the module for identifying ECMO Pointers for the new architecture. This requires additional  engineering efforts and  algorithm~\ref{alg:match_new} is provided.
 \section{Related Work}

\smallskip
\noindent \textbf{Static Firmware Analysis}\tab
Researchers apply the static analysis technique
to analyze the embedded firmware. For instance, 
Costin et al.~\cite{UsenixSec14_Costin} conduct a
large-scale analysis towards the embedded firmware. By analyzing 32 thousand firmware images, many new vulnerabilities are discovered, influencing 123 products. 

Code similarity is widely
used to study the security issue of embedded devices. Feng et al. propose Genius~\cite{feng2016scalable}, a new bug search system to address the scalability issues by translating binary control flow graph to high-level numeric feature vectors. The experiments show that Genius can identify many vulnerabilities in a short time. Considering the inaccuracy of approximate graph-matching algorithm, Xu et al. utilize neural network-based approach to abstract the control flow graph of binary function and build a prototype named Gemini~\cite{xu2017neural}. The result shows Gemini can identify more vulnerable firmware images compared with Genius. Yaniv et al. introduce a precise and scalable tool named  Firmup~\cite{firmup2018} by considering the relationship between procedures. The result show Firmup has a relatively low false positive and effective on discovering vulnerabilities. In the case that firmware images are not available, Xueqiang et al.~\cite{wang2019looking} applies cross analysis of mobile apps to detect  the vulnerable devices. Finally, 324 devices from 73 different vendors are discovered.
Our system is used to analyze the
firmware images of embedded systems with dynamic analysis. Application building upon ECMO can complement
the static analysis ones.

\smallskip
\noindent \textbf{Dynamic Firmware Analysis}\tab
Besides static analysis, researchers propose several methods to support the dynamic firmware analysis.
Avatar~\cite{zaddach2014avatar} is proposed to support complex dynamic analysis
of embedded devices by orchestrating the execution of an emulator and real hardware.
Charm~\cite{talebi2018charm} applies a similar strategy. It introduces
the technique named remote device driver execution by forwarding the
MMIO operation to a real mobile.
Avatar2~\cite{avatar2} extends Avatar to support
replay without real devices. However, they both suffer from the problem of scalability. Inception~\cite{corteggiani2018inception} applies symbolic execution based on KLEE~\cite{cadar2008klee} and a custom JTAG to improve testing embedded software. However, it assumes that the source code is available. IoTFuzzer~\cite{chen2018iotfuzzer} aims to fuzz the firmware from the mobile side. However, the code coverage of firmware and the coverage of attack surface are limited. Pretender~\cite{eric2019pretender} is able to conduct automatically rehosting tasks. However, it replies on the debug interface of specific devices. Jetset~\cite{johnson2021jetset} utilizes the symbolic execution to infer the return values of device registers.
However, the functionality of the peripherals cannot be guaranteed. Furthermore, the shell may not be obtained for further development of different applications.

Besides, many researchers utilize the fuzzing technique to detect the security issues
of embedded firmware.
P2IM~\cite{feng2019p2im} is proposed to learn the model of peripherals automatically. DICE~\cite{dice} focused on the DMA controller and can extend the P2IM's analysis coverage.
Halucinator~\cite{clements2019hal}  proposed a new methodology to rehost the
firmware by abstracting the HAL functions. 
ECMO are different from them in the aspects to transplant peripherals into
the target kernel, instead of inferring the peripherals models. Besides,
all these systems focus on bare-metal system, which is less complicated
than the Linux kernel. 
Firmadyne~\cite{chen2016firmadyne} and FirmAE~\cite{kim2020firmae} target on Linux-based firmware. However, 
they focus on the user-space program, instead of the Linux kernel.

\smallskip
\noindent \textbf{Applications based on QEMU}\tab
There are many applications based on QEMU. For example,
researchers have developed new fuzzing systems~\cite{maier2019unicorefuzz,zheng2019firmafl,triforceafl} based on
QEMU. KVM leverages the device emulation provided by QEMU or the virtio~\cite{russell2008virtio} framework for device virtualization.
The idea of virtio is similar to \sysname. However, virtio requires to change the source code of guest while \sysname works towards the Linux kernel binary.
Virtual machine introspection tools~\cite{garfinkel2003virtual,wang2008countering, dolan2011virtuoso,dovgalyuk2017qemu,fu2013bridging,bahram2010dksm}, which are
helpful for debugging or forensic analysis, utilize QEMU to introspect
the system states. Furthermore, dynamic analysis frameworks
use QEMU to analyze malware
behavior~\cite{egele2007dynamic,moser2007exploring,yan2012v2e,riley2009multi,yin2007panorama,yan2012droidscope}. 
\sysname provides the capability to rehost Linux kernels, which
lays the foundation for apply these applications on embedded Linux kernels.

  \section{Conclusion}
In this work, we propose a novel technique named
{peripheral transplantation} to rehost the Linux kernel of
embedded devices in QEMU.
This lays the foundation for applications that
rely on the capability of runtime state introspection.
We have implemented this technique inside a prototype system
called \sysname and applied it to $815$ firmware images, which consist of 20 kernel versions and 37 device models. \sysname can successfully transplant peripherals for Linux kernels in all images.
Among them, $710$ kernels can be successfully rehosted, i.e.,
launching the user-space shell ($87.1\%$ success rate). Furthermore, we successfully install one Ethernet device driver (i.e., smc91x) on all the rehosted Linux kernels to demonstrate the capability of \sysname to support more peripherals.
We further build three applications to show the usage
scenarios of  \sysname.

%

  \bibliographystyle{plain}
  \bibliography{main}

\end{document}